\documentclass[12pt,preprint]{aastex}

\shortauthors{Chan et al.}

\begin{document}
\title{Could the compact remnant of SN 1987A be a quark star?}

\author{  T.~C.~Chan$^1$, K.~S.~Cheng$^1$, T.~Harko$^1$, H.~K. Lau$^2$, L.~M. Lin$^2$, W.~M. Suen$^3$ and 
X.~L. Tian$^1$  }
\affil{$^1$Department of Physics and Center for
Theoretical and Computational Physics,
 The University of Hong Kong, Hong Kong, P. R. China\\
 $^2$ Department of Physics and Institute of Theoretical Physics,
The Chinese University of Hong Kong, Hong Kong, P. R.
China \\
$^3$McDonnell Center for the Space Sciences, Department of Physics,
Washington University, St. Louis, USA }

\begin{abstract}
The standard model for Type II supernovae explosion, confirmed by
the detection of the neutrinos emitted during the supernova
explosion, predicts the formation of a compact object, usually
assumed to be a neutron star. However, the lack of the detection of
a neutron star or pulsar formed in the SN 1987A still remains an
unsolved mystery. In this paper we suggest that the newly formed
neutron star at the center of SN1987A may undergo a phase transition
after the neutrino trapping time scale ($\sim $10 s). Consequently
the compact remnant of SN 1987A may be a strange quark star, which
has a softer equation of state than that of neutron star matter.
Such a phase transition can induce the stellar collapse and result
in a large amplitude stellar oscillations. We use a three
dimensional Newtonian hydrodynamic code to study the time evolution
of the temperature and density at the neutrinosphere. Extremely
intense pulsating neutrino fluxes, with submillisecond period and
with neutrino energy ($> 30$ MeV) can be emitted because the
oscillations of the temperature and density are out of phase almost
$180^{\circ }$. If this is true we predict that the current X-ray
emission from the compact remnant of SN 1987A will be lower than
$10^{34}$erg s$^{-1}$, and it should be a thermal bremsstrahlung
spectrum for a bare strange star with surface temperature of around
$\sim 10^7$ K.

\end{abstract}


\section{Introduction}

Supernova 1987A, the brightest supernova seen in modern times,
was a milestone in astronomy and fundamental physics. The
detection of the neutrinos from SN 1987A \citep{Bi87, Hi87},
confirmed the basic predictions of the physical mechanisms for
Type II supernovae. The explosion is triggered by the collapse of
massive stars. The core collapse of the
massive star (with mass greater than $10M_{\odot}$) is due to
electron capture and photodisintegration, and it is halted when
the center density exceeds the nuclear matter density. As the stellar core collapses, the 
gravitational energy is released through the emission of neutrinos.
A small fraction of these neutrinos is absorbed, and the heating of
the neutrinos drives the supernova explosion.

The supernova SN 1987A was observed in every band of the
electromagnetic spectrum, from radio to gamma rays, and hence it
has a well-measured bolometric light curve. By using both the
bolometric light curve and the spectral evolution of its H$\alpha
$ line, the hydrodynamic and time-dependent atmosphere models are
consistent with a presupernova radius of 35$\pm 5R_{\odot}$, an
ejecta mass of $18\pm 1.5M_{\odot}$, an explosion energy of
$\left( 1.50\pm 0.12\right) \times 10^{51}$ erg, and a radioactive
$^{56}$Ni mass of $0.0765M_{\odot}$ \citep{Ut07}. The inferred
energy ($\sim 3\times 10^{53}$ ergs), temperature ($\sim 10$ MeV)
and decay time ($\sim 4$ s) of the first neutrino burst were not inconsistent with what
would be expected from the production of a neutron star through
core collapse \citep{Hi87}. Presently, about $35$ plausible, or at least
possible, associations of pulsars with supernova
remnants are known, and these are all relatively young pulsars, mostly with $%
\tau _{p}=P/2\dot{P}<10^{5}$ years, where $\dot{P}$ is the first
derivative of the pulsar period $P$ \citep{Ma07}. Almost
immediately after the observation of SN 1987A, optical attempts
were made to identify the compact remnant of the explosion (see
\citet{Ma07} for a recent review of the optical and X-ray
observations on the supernova). The optical searches
were done by \citet{Pe95}, using the High Speed Photometer System
on the Hubble Space Telescope, and by \citet{Ma96}, using the 3.9
m Anglo-Australian Telescope, respectively. No significant
pulsations were observed in the period range of $0.2$ ms to $10$
s, with an upper limit for the pulsed emission equivalent to a V
magnitude of about $25$. However, most pulsars are detected at
radio wavelengths. An extensive search was carried out recently at
the Parkes $64$ m radio telescope in 2006, at frequencies between
$1390$ MHz and $8370$ MHz, respectively \citep{Ma07}. No
significant candidate with a signal to noise ratio greater than
$9.0$ was observed, and there is no observational evidence for the
presence of a central point (or near point) source at any
wavelength \citep{Pa04,Gr05,Sh05}. The limit on the luminosity of
any point source at the center of SN 1987A
is of the order of $10^{33}$ erg s$^{-1}$- $10^{34}$ erg s$^{-1}$ \citep{Ma07}
.

Therefore, the question of why no compact object is observed at
the center of the SN 1987A explosion is a rather intriguing one.
Several explanations have been advanced to explain the lack of a
neutron star/ pulsar. The most obvious one is that the pulsar may
not be beamed towards us \citep{Ma07}. The beaming fraction
(fraction of the celestial sphere swept over by the beam as the
star rotates) is not very well determined, but in radio band a
beaming fraction of $20\%$ is usually assumed, and for young
pulsars it could be even larger \citep{Ma07}. However, even if the
pulsar were not directed along our line of sight, its radiation
would heat the surrounding supernova remnant, and add to its
bolometric luminosity. At present, all of the luminosity of SN
1987A can be accounted for by a radioactive decay model
consistent with the production of $0.075M_{\odot}$ of $^{56}$Ni
\citep{Fr99, Ut07}. It would be also possible that the pulsar
magnetic field develops in a few decades, and in this case a
rapidly spinning neutron star would be still undetectable
\citep{Ma07}. However, most of the current models assume that the
magnetic field is either frozen in the star, or it is generated
during the collapse by the dynamo process \citep{Bo05}. Another
possibility is that the neutron star further collapsed into a
black hole, whose accretion luminosity is below the observational
limits \citep{Be95}. Although the duration of the neutrino burst
of SN 1987A precludes further collapse in the first $20$ s, the
neutron star could have collapsed eventually into a black hole
because of either accretion of a sufficient fall-back mass, or due
to some changes in the equation of state of the dense matter
\citep{Fr99}. The analysis of the collapse of the initially formed
neutron star in SN 1987A puts severe constraints on the equation
of state of the dense matter. However, to explain the collapse of
a low mass neutron star an exotic equation of state of the nuclear
matter is needed, like, for example, an equation of state softened
by pion condensation \citep{Be95}, which would contradict other
observations on supernova remnants. Initially, it is likely that
neutron stars formed in Type II supernova explosions are obscured
by the late-time fallback. Much of this fallback is quickly
accreted via neutrino cooling, but some material may remain on the
neutron star, forming an atmosphere that slowly accretes through
photon emission. If the neutron star has either a low magnetic
field, or a low rotational spin frequency, then the neutron star
remnant of SN 1987A can not be seen \citep{Fr99}.

When studying the neutrino data from the SN 1987A explosion a
number of unexpected features have been found \citep{Hi87,Bi87,Co04,Co04a}.
For example, the angular distribution of the two events, one seen
at the Kamiokande-II (KII) \citep{Hi87} and the other at
Irvine-Michigan-Brookhaven (IMB) \citep{Bi87} are more forward directed than
expected. The average cosines of the polar angles for these two
events are $\left<\cos \theta ^{KII}\right>\approx0.3$, while
$\left<\cos \theta ^{IMB}\right>\approx0.5$. Moreover, the energy
distribution of these two detectors are not in perfect agreement.
The mean energy of the neutrinos detected at KII is around half of
the energy of the neutrinos detected at those detected at IMB,
with a mean energy $\left<E^{IMB}\right>\approx 30$ MeV. The time
distribution of the two events is also very different, on average the neutrinos detected
by IMB came $\sim 5$ s after those detected by KII (see \citet{Al88} for a summary of
neutrino arrival time for various detectors). When fitted
with thermal spectra, the two independent detections do not seem
to agree with either each other or typical theoretical
expectations. Using parameter-free inferential statistical methods, it can be shown that
the combined KII and IMB data can be best explained by a spectral shape that is enhanced
both at the peak and the tail of the spectrum, and depressed in between, as compared to
the Fermi-Dirac spectrum  \citep{Yu07}. Since these methods make no {\it a priori}
assumptions and do not rely on parameter estimation, they allow for more efficient
processing of small data samples. While the supernova neutrino spectra are expected to be
quasi-thermal, modifications due to non-standard effects, like neutrino mixing among
various flavors, neutrino decay, neutrino-neutrino interactions and/or any novel
mechanism due to unknown physics may produce a time-integrated spectrum that deviates
significantly from a quasi-thermal shape \citep{Yu07}.

It is the purpose of the present paper to propose some
explanations for the lack of observational evidence for the
presence of a point source at the center of SN1987 A, as well as the apparent
discrepancy between the two neutrino detections.

Since \citet{Wi84}, following early proposals by \citet{It70} and
\citet{Bo71}, suggested that strange quark matter, consisting of
$u$-, $d$- and $s$-quarks is energetically the most favorable
state of the matter, the problem of the existence of strange quark
stars has been intensively investigated in the physical and
astrophysical literature. The possibility that some compact
objects could be strange stars remains an interesting and
intriguing, but still open question. \citet{Wi84} also proposed
two ways of formation of strange matter: the quark-hadron phase
transition in the early universe and conversion of neutron stars
into strange ones at ultrahigh densities. In the theories of
strong interaction quark bag models suppose that breaking of
physical vacuum takes place inside hadrons. As a result vacuum
energy densities inside and outside a hadron become essentially
different, and the vacuum pressure on the bag wall equilibrates
the pressure of quarks, thus stabilizing the system. If the
hypothesis of the quark matter is true, then some of the neutron
stars could actually be strange stars, built entirely of strange
matter \citep{Ha86,AlFa86,AlOl89}. However, there are general
arguments against the existence of strange stars, e. g. \citet{Ca91}. For
a general review of strange star properties and the physics of
phase transitions see \citet{Ch98} and \citet{Ha07}.

A possibility for the formation of the quark stars is that some
neutron stars in low-mass X-ray binaries can accrete sufficient
mass to undergo a phase transition to become strange stars
\citep{Ch96}. This mechanism has
also been proposed as a source of radiation emission for cosmological
$\gamma$-ray bursts, soft gamma-ray repeaters or other
astrophysical objects \citep{Ch98a,ChDa98,Pa05}. Some basic
properties of strange stars like mass, radius, cooling, collapse
and surface radiation have been also studied \citep{Ch00, Ch03,
Ha00, Ha02,Zd01,Go03,Be04,Zd06}. Quark stars are expected to form
during the collapse of the core of a massive star after the
supernova explosion, as a result of a first or second order phase
transition, resulting in deconfined quark matter \citep{Ta85,
Ta86, Ge93, Da95}. The proto-neutron star core, or the
neutron star core, is a favorable environment for the conversion
of ordinary matter to strange quark matter \citep{Chen07}.

This paper is organized as follows. In Section 2, we discuss various
possible phase transition processes in a neutron star, and to
estimate the characteristic time scale for the phase transition. We
suggest that the phase transition from neutron star matter to
strange matter in the core of neutron star may be more favorable. In
Section 3, we present the numerical simulation results of a
phase-induced collapse neutron star. In particular, we show why high
intensity and high energy neutrinos can be emitted after the phase
transition. In Section 4, we summarize the energy dissipation
processes of bare strange stars and their cooling. In Section 5, we
give a brief discussion about how the phase transition may take
place in the compact remnant of 1987A and final remarks are
presented.

\section{Phase transitions in high density neutron matter}\label{sect2}

During the formation and evolution of neutron stars new states of
matter may form inside the stars as a result of a phase transition,
which may be triggered by the accretion of matter, pulsar spindown,
collapse of the core of a proto-neutron star in supernova explosions
etc. In the following we shall restrict our discussion only to phase
transitions of the first order. As a general physical model we will
discuss the nucleation of a phase $B$ in the metastable phase $A$.
The nucleation is concerned with fluctuations of parameters, such as
the local density or the number of particles in a metastable drop of
phase $B$ which triggers the phase transition \citep{LaLi80}. There
are two extreme cases of the nucleation theory. In the classical
regime, the temperature is assumed to be sufficiently high to
trigger the phase transition by thermal fluctuation. In the quantum
regime, below a characteristic temperature, thermal fluctuations are
negligible as compared to the quantum ones. Hence quantum
fluctuations can initiate a phase transition via the quantum
tunneling effect. The relevant thermodynamic potential for the
description of phase transitions is the Gibbs free energy
\citep{HaSc82}. In the following we denote by $\Delta F$ the excess
free energy of a critical droplet.

The probability of a local formation of a droplet of phase $B$ via a
fluctuation decreases very strongly with the increase of the number of
baryons $A_{drop}$ in the droplet. For small $A_{drop}$, however, the
positive contribution of the surface energy to $\Delta F$ prevails over the
gain in the bulk binding, $\Delta F\left( A_{drop}\right) >0$, which makes
the droplet unstable with respect to the reconversion to phase $A$. However,
at some value $A_{drop}=A_{crit}$, the energy excess due to the droplet
formation vanishes, $\Delta F\left( A_{crit}\right) =0$. Therefore the
droplets with $A_{drop}>A_{crit}$ grow spontaneously, destabilizing the
metastable phase $A$ and inducing the phase transition \citep{Ha04}.

By using the thermodynamical formalism of nucleation \citep{LaLi80,HaSc82,Ha07} one can
study the nucleation of exotic phases (pion condensate, kaon condensate and
quark matter) in the high density neutron matter. In the following we will
briefly review each of these processes.

\subsection{Meson condensation}\label{sect21}

The nucleation of the pion condensate in the neutron star core and its astrophysical 
implications were studied
by \citet{HaSc82,HaPr82,MuTa90}, respectively, where a
theory of metastability of dense neutron matter with respect to the
first-order phase transition to a pion-condensed state was also analyzed.
The description of the condensation process is based on the idea of the
nucleation of the pion-condensed phase in the metastable normal matter,
through the appearance of spontaneously growing droplets of the new phase.
Different paths leading from the false to the true ground state of dense
neutron matter have been considered in \citet{HaSc82}.

The calculations performed for realistic models of cold neutron matter yield
an interval of metastability (on the time scale of the age of the universe),
which is as large as half of that between the baryon density $\rho _{N}$,
where the metastability starts, and $\rho _{crit}$, where the potential
barrier (without surface effects) between the true and the false ground
state vanishes \citep{HaSc82}.  In the case of the pion condensation in hot
neutron matter ($T\sim 5$ MeV) the region of metastability (on the time
scale of the gravitational collapse of a massive star) is much narrower than
in cold neutron matter. Consequently, additional energy release and entropy
generation from the first-order phase transition in hot supercompressed
matter are  negligible. The value of the $A_{crit}$ strongly decreases with
the growth of the overcompression $\Delta P_{over}=P-P_{0}$. Thus, the
lifetime of an overcompressed state decreases rapidly with increasing $%
\Delta P_{over}$ \citep{HaSc82,Ha07}. For the case of the pion condensation the
surface tension can be approximated by the expression derived by \citet{Ba71a}%
, with the nucleus replaced by a pion condensed droplet, and the neutron gas
replaced by the ordinary neutron star matter. Then, one can obtain the
condition for the pion condensate in the neutron star core during a time
interval equal to the present age of the universe, $t_{H}=1.5\times 10^{10}$
years. Depending on the employed model, the required overcompression is $%
\Delta P_{over}/P_{0}=0.02$ or $\Delta P_{over}/P_{0}=0.05$ \citep{MuTa90,Ha07}.

Under typical conditions in the neutron star core ($T\leq 10^{9}$ K) the
nucleation proceeds via the quantum tunneling through the energy barrier. At
much higher temperatures, the thermal effects increase the nucleation rate
through thermally excited droplet states. On the other hand, the growth of
the temperature increases $P_{crit}$. For a newly born neutron star with $%
T\geq 10^{10}$ K, the nucleation proceeds in the classical (thermal) regime \citep{Ha07}.
Due to the softening of the equation of state, pion condensation could lead
to a significant decrease of the maximum mass and moment of inertia
allowable for neutron star models, constructed using such an equation of
state \citep{HaPr82}.

The formation of a droplet of kaon condensate in a neutron star core is
connected with the production of strangeness, and nucleation should involve
the weak interaction process. The transition from $npe$-type nuclear matter
(consisting of neutrons, protons, and electrons) to matter containing
strangeness, using a Walecka-type model, predicting a first-order
kaon-condensate phase transition, was studied in \citet{No02}. The free
energy of the droplets of the kaon-condensed matter, as well as the density, temperature,
and the neutrino fraction were obtained.
The surface tension of the interface between the normal and condensed kaon phase was
calculated by \citet{Ch00}, using the non-uniform relativistic mean field
model. In the approximation in which only linear terms in the curvature of
the surface are kept, the surface contribution to the thermodynamic
potential of a spherical droplet is $\sigma =\sigma _{S}+2\sigma
_{c}/R_{drop}$, where $\sigma _{S}$ is the surface tension and $\sigma _{c}$
is the curvature coefficient. For a small admixture of kaon-condensed
droplets in a nucleon matter $\sigma _{S}=30$ MeV \citep{Ch00}.

The case of the kaon condensation is drastically different from the case of
the deconfinement transition, since here there is no intermediate
zero-strangeness state, which might allow for the fast nucleation followed
by a slow but smooth growth of the strangeness containing fields. Instead,
the thermal fluctuations responsible for nucleation events must directly
involve the weak-interaction processes which produce kaons. The weak
interaction processes producing strangeness via the reaction $e+N\rightarrow
\nu _{e}+K^{-}+N$ (where an additional nucleon $N$ is needed for momentum
conservation) and $n\rightarrow p+K^{-}$ are too slow to create a critical
droplet of kaon condensate from a density fluctuation during the fluctuation
lifetime \citep{No02}. Strangeness can be produced at a reasonable rate from thermal
kaon-antikaon ($K^{-}K^{+}$) pairs, but this mechanism can operate only at
extremely high temperatures typical for protoneutron stars. Generally, a
protoneutron star cools so rapidly that kaon condensate has no time to
nucleate \citep{Ha07}. However, as soon as $\mu _{e}>\omega _{K^{-}}^{0}$, where $\omega
_{K^{-}}^{0}$ is the minimum energy of a single zero momentum kaon in dense
matter, a spontaneous formation of kaons is possible \citep{Ha07}. The kinetics of kaon
condensation was studied by \citet{Mu97}, \citet{Mu00a} and \citet{Mu00b}.
Kaon potentials $U_K$ with values only a little below $U_K=-120$ MeV would not be compatible with 
the mass of the Hulse-Taylor pulsar, because of a mechanical instability that is 
initiated by the central densities for which the pressure remains constant, and the 
necessary condition for stability $dM/d\rho _c>0$ is not satisfied \citep{Gl00}. The kaon 
condensation process in neutron stars in the framework of the Zimanyi-Moszkowski model in 
the relativistic mean field theory was considered in \citet{DaCh97}. 
Even though hyperons which may increase 
the critical condensation density are not included, kaon condensation may not occur in 
stable neutron stars for these classes of mean field theories. The existence of the 
antikaon condensation phase in neutron stars in the frameworks of the 
Glendenning-Moszkowski and Zimanyi-Moszkowski models was reanalyzed in \citet{Wan07}. The 
results of this analysis show that in the very massive ($M=2\pm 0.2M_{\odot}$), and high 
redshift neutron stars, there are still some stiff enough equations of state of neutron 
matter, so that the pure antikaon condensation phase and the mixed phase of normal 
baryons and antikaon condensation can still exist. The mass of the neutron star inferred 
from the neutrino flux of SN 1987A should be of the order of $1.5-1.7M_{\odot }$, which 
suggest that the kaon condensate may not occur in the compact object of the SN1987A.

\subsection{Quark deconfinement}\label{sect22}

When quarks seeds are formed in the core of the neutron stars, they
will propagate through the entire star, and convert it to the new phase.
A phase transition occurs between the hadronic and quark phase when the
pressures and the chemical potentials in the two phases are equal,
$P_{h}=P_{q},\mu _{h}=\mu _{q}$, where $P_{h}$, $\mu _{h}$ and $P_{q}$, $\mu _{q}$ are 
the pressures and the chemical potentials in the hadron and quark phase, respectively. If 
the transition pressure is less than that existing in the
supernova core the transition can occur.

The change from the metastable neutron matter phase to the stable
quark phase occurs as the result of fluctuations in a homogeneous medium,
formed of neutrons, in which small quantities of the quark phase (called
bubbles or nuclei) are randomly generated. Since the process of creation of
an interface is energetically unfavorable, it follows that when a quark
nucleus is below a certain size, it is unstable and disappears again.
Surface effects disfavor the survival of small bubbles below the radius $%
R_{c}$ (called critical size-the nuclei of this size are called critical
nuclei or bubbles), which is nothing but the value that extremizes the
thermodynamical work $W$ necessary to create the bubbles \citep{AlOl89, Ha04}. Only nuclei whose
size $r$ is above the value $R_{c}$ are stable, and can survive %
\citep{LaLi80}. The nuclei are assumed to be macroscopic objects containing
a large number of particles (quarks).
Following the phase transition to the two flavor quark matter, the two flavor quark matter will convert into three
flavor quark matter through the reactions $u+e^{-}\leftrightarrow d+\nu _{e}$
, $u+e^{-}\leftrightarrow s+\nu _{e}$, and $u+d\leftrightarrow u+s$,
respectively \citep{Da95}.

Once the quark phase is formed inside the neutron star, it will
propagate throughout the entire star. The physical mechanisms of the
transition from neutron matter to quark matter in an astrophysical
background have been studied within several models. The first is due to \cite
{Ol87}, who used a non-relativistic diffusion model. As such, this is a slow
combustion model, with the burning front propagating at a speed of
approximately $10$ m/$\sec $. This is determined primarily by the rate at
which one of the down quarks inside the neutrons is converted, through a
weak decay, to a strange quark: $d+u\rightarrow s+u$. The second method of
describing the conversion process was first suggested by \citet{Ho88}, and
analyzed in detail by \citet{Lu94} and \citet{Lu95}, who modeled the
conversion as a detonation. In this case the conversion rate is several
orders of magnitude faster than that predicted by the slow combustion model.
This model is based on the relativistic shock waves and combustion theory.
But regardless of the way in which the transformation occurs, an initial
seed of quark matter is needed to start the process.

The problem of the combustion of neutron matter is closely related to that
of shock waves. Let us assume that we have an unburnt fluid which converts
(by means of a certain reaction) in a burnt fluid. The combustion process
must be exothermic if it propagates spontaneously to other regions of the
fluid. The condition for spontaneous propagation to other region of the
fluid is $E_{burnt}\left( P,X\right) <E_{unburnt}\left( P,X\right) $, where $%
E_{burnt}\left( P,X\right) $ and $E_{unburnt}\left( P,X\right) $ are the
energy densities of the respective fluids, both evaluated at the same
thermodynamic state. In the case of the transition from nuclear to quark
matter this condition can be reformulated as $E_{nucl}-3P_{nucl}>4B$ \citep{Lu94},
where $E_{nucl}$ and $P_{nucl}$ correspond to nuclear matter. If
this condition is not fulfilled, the combustion is no longer
exothermic, and so it is not possible. At low enough densities, soft
EOS verify $E_{nucl}>>P_{nucl} $, and $E_{nucl}\approx
m_{n}n_{B}c^{2}$. Hence there exists an absolute lower limit for the
combustion to be possible, given by
$n_{B}\approx 4B/m_{n}c^{2}=0.25\left[ B/(145
\mathrm{MeV})^4\right] \mathrm{fm}^{-3}$ \citep{Lu94},
corresponding to a transition density $\rho _{tr}$ of the order of
$\rho _{tr}\approx 4.2\times 10^{14}\left[
B/(145\mathrm{MeV})^4\right] \;\mathrm{g/cm^{3}}$.

However, it is important to point out that the efficient burning of
the nuclear matter can take place only when the nuclear matter
density is sufficiently high, and in order to convert the entire
neutron star to a quark star a higher density than the minimum one
given by $n_B$ is required, and there is always a
range of densities at which the transition occurs. The density for
the deconfinement of baryonic matter with a moderately stiff (or
stiff) equation of state to two-flavor quark matter is near $8\rho
_{nuc}$. For a soft equation of state, the deconfinement density may
be lower.

Generally, the flow behind the detonation is sonic. Therefore the typical
time scale for the transition is $\tau _{tr}=R/c_{s}$, where $R$ is the
radius of the neutron star and $c_{s}$ is the speed of the sound. A simple
phenomenological model for the evolution of the quark phase can be obtained
by assuming  $dr/dt=\left(r-R_{c}\right)/\tau _{tr}$ \citep{Ha04},
which gives for the transition time scale $T_{tr}$ from a microscopic quark nugget to a
quark matter distribution of a macroscopic size the expression
\begin{equation}
T_{tr}\approx 10^{-4}N_q^{-1/3}R_6\ln \frac{R}{ R_{q}}~{\rm s},
\end{equation}
where $R_{6}$ is the neutron star radius in units of $10^6$ cm, $R_q
\sim 300R_c$ (Harko et al. 2004) is the initial size of the quark
bubble, and $N_q$ is the number of quark seeds inside the core of
the neutron star, which could be as large as $10^{48}$ (Iida \&
Sato 1998). However, it was found that the conversion process in a
hadronic matter at $T=0$ always correspond to a deflagration, and
never to a detonation \citep{Dr07}. Hydrodynamical instabilities can
develop on the front, and a  mixed phase of hadrons and quarks could
form. Due to the formation of the wrinkles, the conversion velocity
can significantly be increased, but this increase is not sufficient
to transform the deflagration into a detonation. In general, one
could assume that the conversion process can take place in two
steps, with a first transition from hadrons to ungapped (or 2SC)
quarks, followed by a second transition in which a CFL phase is
produced. The time scale for the first transition has been estimated
by \citet{Dr07}, and it is of the order of $0.1-1$ s for the first
transition in the case of a laminar front, and is much more rapid if
the hydrodynamical instabilities are taken into account. The second
transition lasts only $10^{-3}$ s, due to the formation of a
convective layer. If the two processes take place one after the
other, it is even possible that the formation of a diquark
condensate could accelerate the conversion process by developing a
convective layer inside the hadronic phase. The neutrino trapping
can play an important role in the hadron - quark phase transition
\citep{Vi05}. The quantum nucleation of a quark matter drop, and
therefore the conversion of the hadronic matter to quark matter, is
strongly inhibited at $T=0$ by the presence of neutrinos. However,
in the high temperature regime, the dominant nucleation mechanism is
thermal nucleation, and not quantum tunneling \citep{Ha07}. In
general, it takes a sub-milliseconds time interval to convert the
matter at the core of a neutron  core into quark matter.

\section{Neutrino emission from a phase-induced collapse neutron star}

The total energy emitted in the form of neutrinos in a supernova
explosion can be easily estimated from qualitative considerations.
The total gravitational energy that can be irradiated is
$E_{b}\approx 3GM^{2}/5R$, where $M$ and $R$ are the mass and the
radius of the star, respectively. By using for the neutron star a
mass of the order of $M=(1-2)M_{\odot}$, and a radius of $R=20$ km
$\left( M_{\odot}/M\right) ^{1/3}$, one obtains $E_{b}\approx \left(
1-5\right) \times 10^{53}$ erg. This amount of neutrino energy does
not conflict with the observed data. However, it is difficult to
understand why in comparing with KII,  IMB has detected a neutrino
burst with a time delay ($\sim 5 s$) and with even higher average
energy ($\sim 30$MeV). In discussion section, we will argue that a
phase transition may take place after the core temperature of the
newly born neutron star is reduced. The characteristic time scale
for the cooling of the core is the neutrino trapping time, which is
of the order of several seconds. In this section, we use a three
dimensional Newtonian hydrodynamic code to simulate the neutrino
emission from a phase-induced collapse of neutron star.

\subsection{Description of numerical code}

First we briefly summarize the numerical code used to simulate the
collapse of neutron star induced by phase-transition. The
three-dimensional simulations are based on Newtonian hydrodynamics
and gravity. The code has been used to study the gravitational wave
emission from the phase-induced collapse neutron stars (Lin et al.
2006). We refer the reader to Lin et al.(2006) for a detailed
discussion of the numerical code.

The system of equations describing the non-viscous Newtonian
fluid flow is given by
\begin{equation}
        \frac{\partial \rho}{\partial t} + \nabla \cdot
        \left( \rho {\bf v} \right) = 0 ,
\label{eq:rhoeq}
\end{equation}
\begin{equation}
        \frac{\partial}{\partial t} \left( \rho v_{i} \right) +
        \nabla \cdot \left( \rho v_{i} {\bf v} \right) +
        \frac{\partial P}{\partial x_{i}} = - \rho
        \frac{\partial \Phi}{\partial x_{i}} ,
\label{eq:momeq}
\end{equation}
\begin{equation}
        \frac{\partial \tau}{\partial t}
        + \nabla \cdot \left(
        \left( \tau + P \right)
        {\bf v} \right) = -\rho {\bf v} \cdot \nabla \Phi  ,
\label{eq:taueq}
\end{equation}
where $\rho$ is the mass density of the fluid, ${\bf v}$ is the
velocity with Cartesian components $v_i$ ($i=1,2,3$), $P$ is the
fluid pressure, $\Phi$ is the Newtonian potential and $\tau$ is the
total energy density, $\tau = \rho \epsilon + \rho {\bf v}^{2}/2$ ,
and $\epsilon$ is the internal energy per unit mass of the fluid.
The Newtonian potential $\Phi$ is obtained by solving the Poisson
equation, ${\nabla}^{2} \Phi = 4 \pi G \rho $. The system is
completed by specifying an equation of state $P=P(\rho, \epsilon)$.

The above hydrodynamics equations (\ref{eq:rhoeq})-(\ref{eq:taueq})
can be rewritten in a flux conservative form, which can be solved
numerically using quite standard high-resolution shock capturing
(HRSC) schemes. An HRSC scheme has the ability to resolve
discontinuities (e.g. shock waves) in the solution. It can also
achieve high accuracy in regions where the fluid flow is smooth. In
our work, we employ the so-called Roe's solver in the simulations
(see Lin et al. 2006).

It is  not known what is the equation of state (EOS) for the neutron
star in the remnant of supernova 1987A. We can try all possible
existing realistic equations of state in our study. However, the
main purpose of this paper is to demonstrate that the phase-induced
collapse of a neutron star can emit extremely intense, pulsating and
very high energy neutrinos. Therefore instead of using realistic
equation of state we will use a polytropic equation of state for the
initial neutron star. In Section 2, we have discussed various
possible phase transitions for a newly born hot neutron star.
Although both pion condensate and kaon condensate cannot be ruled
out completely, there are no compelling observation evidence for the
their existence. For simplicity we will focus our study on the phase
transition from neutron star to strange star in this paper. We use a
mixed phase EOS to mimic a strange star covered by normal nuclear
matter.

In the following we want to argue that a newly born quark star
should be described by a mixed phase EOS. In section 2.2, it has
been shown that quark seeds can be spontaneously generated inside
the core of neutron star when the density exceed the critical
density (e.g. Iida \& Sato 1998). These quark seeds formed in the
neutron star core can mix with the normal hadronic matter by
Schwarzschild convection, and the length scale $\lambda _{c}$ for
the convective motion is given by \citep{Wi88a}
\begin{equation}
\lambda _{c}=\left[ \frac{k_{T}\eta }{g\left( -d\ln S/dr\right) }\right]
^{1/4},
\end{equation}
where $S$ is the entropy per baryon in units of Boltzmann's constant, $g$ is
the local gravitational acceleration in the core, $k_{T}=\lambda _{\nu }/3$,
where $\lambda _{\nu }$ is the neutrino mean free path, $\eta =\lambda _{\nu
}\rho _{\nu }/3\rho $, where $\rho _{\nu }$ is the neutrino energy density
and $\rho $ the matter density. With $\lambda _{\nu }=80$ cm, $\rho _{\nu
}/\rho =0.1$, $g=8\times 10^{13}$ cm/s$^{2}$ and $-d\ln S/dr=5\times 10^{-6}$
cm$^{-1}$\citep{Da95}, one obtains $\lambda _{c}=3.5\times 10^{3}$ cm. The
timescale of the mixing can be calculated from $\tau _{c}=2\pi /\sqrt{%
g\left( -d\ln S/dr\right) }=0.3$ ms. Therefore the convection will
result in the mixing of the neutron and quark phases in the large
region of the star during the phase transition process \citep{Gl00}.

The initial equilibrium neutron star before the phase is given by a polytropic EOS
\begin{equation}
P=k_0\rho^{\Gamma_0} ,
\label{eq:poly_EOS}
\end{equation}
where $k_0$ and $\Gamma_0$ are constants.
On the initial time slice, we also need to specify the specific internal
energy $\epsilon$. For the polytropic EOS, the thermodynamically consistent
$\epsilon$ is given by
\begin{equation}
\epsilon = {k_0\over \Gamma_0-1} \rho^{\Gamma_0-1} .
\end{equation}
Note that the pressure in Eq.~(\ref{eq:poly_EOS}) can also be written as
\begin{equation}
P=(\Gamma_0-1)\rho\epsilon .
\end{equation}

We assume that the phase transition take place at $t=0$, then we
switch the compact object from a polytropic EOS to the EOS of a
mixed phase quark star, which consists of two parts : (i) a mixed
phase of quark and nuclear matter in the core at density higher than
a certain critical value $\rho_{tr}$ (quark seeds can spontaneously
produce everywhere when $\rho \geq \rho _{tr}$) (ii) a normal
nuclear matter region extending from $\rho < \rho_{tr}$ to the
surface of the star. Explicitly, the pressure is given by
\begin{equation}
P = \left\{ \begin{array}{cc}
           \alpha P_{\rm q} + (1-\alpha) P_{\rm n}
         & \ \mbox{for} \ \  \rho > \rho_{tr} \\
        \\
            P_{\rm n}
         & \mbox{for} \ \ \rho  \leq \rho_{tr} ,  \end{array}  \right.
\label{eq:mixed_EOS}
\end{equation}
where
\begin{equation}
P_{\rm q} = {1\over 3} \left( \rho + \rho\epsilon - 4B\right)
\label{eq:quark_eos}
\end{equation}
is the pressure contribution of the quark matter, and
\begin{equation}
P_{\rm n}=(\Gamma _n-1)\rho \epsilon
\label{eq:eos_idealgas}
\end{equation}
where $\Gamma _n$ is not necessarily equal to $\Gamma _0$, is that of the nuclear matter, 
and
\begin{equation}
\alpha = \left\{ \begin{array}{cc}
          { (\rho -\rho_{tr}) / (\rho_{q} - \rho_{tr}) }
         & \ \mbox{for} \ \  \rho_{tr} < \rho < \rho_{\rm q} \\
        \\
            1
         & \mbox{for} \ \ \rho_{q} < \rho ,  \end{array}  \right.
\label{eq:alpha}
\end{equation}
is defined to be the scale factor of the mixed phase (Lin et al.
2006). We should notice that $P_q$ is  not in the usual form of MIT
bag $P_q = {1\over 3}(\rho_{\rm tot} - 4 B)$, where $\rho_{\rm tot}$
is the (rest frame) total energy density, and  $B$ is the bag
constant. It is because in Newtonian simulation, we use the rest
mass density $\rho$ and specific internal energy $\epsilon$ as
fundamental variables in the hydrodynamics equations. The total
energy density $\rho_{\rm tol}$, which includes the rest mass
contribution, is decomposed as $\rho_{\rm tot} = \rho +
\rho\epsilon$. We choose $\Gamma _n< \Gamma_0$ in our simulations to
take into account the possibility that the nuclear matter may not be
stable during the phase transition process, and hence some quark
seeds could appear inside the nuclear matter, or the convection
mentioned in the early of this section may mix some quark matter
with the nuclear matter. In the presence of the quark seeds in the
nuclear matter, the effective adiabatic index will be reduced. The
possible values of $B^{1/4}$ range from 145 MeV to 190 MeV
\citep{DeGr75, Sa82, Ste95}.
For $\rho >\rho _q$, the quarks will be deconfined from nucleons. The value of $\rho _q$ is
model dependent; it could range from 4 to 8 $\rho_{nuc}$ \citep{Ch98a, Ha03, Bo04},
where $\rho_{nuc}=2.8\times 10^{14}~{\rm g~cm}^{-3}$ is the nuclear
density.

In the simulations, we set $\Gamma_0=2$, $\Gamma_n = 1.85$, $B^{1/4}
= 160$ MeV, and $\rho_q = 9 \rho_{\rm nuc}$. The phase-transition
density $\rho_{tr}\approx 2.1 \rho_{nuc}$ is defined to be the point
where $P_q$ is zero initially. This value is approximately equal to
the value estimated in section 2.2. The total time span of each run
is $\sim$5 ms, the time step is $3.7\times 10^{-4}$ ms. The grid
spacing is set to be $dx=0.28$ km and the outer boundary of the
computational domain is at 27.5 km, which is about two times the
stellar radius of our models. During the numerical simulation, a
low-density atmosphere is added outside the neutron star for
numerical stability purpose. The density and temperature of the
atmosphere are $3 \times 10^{9}\ \textrm{g} / \textrm{cm}^3$ and
$0.003$ MeV, respectively.

Finally we want to remark that although eq.(9) is not the exact
situation of quark matter distribution inside the star, it allows us
to simulate the phase induced collapse and study the energy
transport processes. The exact quark matter distribution can
certainly affect evolution of star in detail and neutrino emission
from the stellar surface quantitatively. However since eq.(9) should
still represent the qualitative quark matter distribution therefore
the simulated features produced by eq.(9) should be still
qualitatively correct.

\subsection{Neutrino luminosity}

For a new born neutron star, the internal temperature is so hot that
neutrinos will be trapped inside the star for at least a few seconds (cf.
\citet{ShTe83} for a general review). However, neutrinos very near
the surface of the star can still escape from the star because the optical
depth near the stellar surface is low. In fact we can define a radius ($R_{\nu}$)
called neutrinosphere, where the optical depth of electron neutrinos is given by
\begin{equation}
\label{eq:Janka2001eq16}
\tau_{eff}=\int_{R_\nu}^\infty dr \,
\kappa_{eff}(r)=1
\end{equation}
where the effective optical depth, $\tau_{eff}$ is defined as
inverse mean free paths and the effective opacity, $\kappa_{eff}$ is
given by \citep{Janka2001}
 \begin{equation}\label{eq:opacityFinal}
\langle\kappa_{eff}\rangle(r)=1.202\times
10^{-7}\rho_{10}(r)\left(\frac{T_\nu(r)}{4\textrm{MeV}}\right)^{2}\quad\frac{1}
{\textrm{cm}},
\end{equation}
where $\rho _{10}(r)$ is the density in units of $10^{10}$ g/cm$^3$, and $T_{\nu}(r)$ is
the temperature at $r$ respectively. The electron neutrino luminosity emitted from the
neutrinosphere is given by \citep{Janka1995, Balantekin2005}
\begin{equation}
L_\nu= \frac{7}{16} \pi R_\nu^2 ac T_\nu^4.
\end{equation}
In our simplified calculation, we also assume equal luminosities for
neutrino and antineutrino, hence the combined luminosity for a
single neutrino flavor is
\begin{eqnarray}\label{eq:lum}
L_{\nu, \, \overline{\nu}} &=& L_\nu + L_{\overline{\nu}}\nonumber\\
&=& \frac{7}{8} \pi R_\nu^2 ac T_\nu^4
\end{eqnarray}

Since the neutrinosphere is very near the stellar surface, where the
nuclear matter is described by an ideal gas EOS $P_n = (\Gamma_n -
1)\rho \epsilon$ (see previous section), we thus approximate the
temperature by $T = 2m_B\epsilon / 3 k $, where $k$ is the
Boltzmann's constant and $m_B=1.67\times 10^{-24}$ g is the baryon
mass. With the simulated density and temperature profiles at a given
time, we obtain the value of each $R_\nu$ from 0 to 3 ms with
trial-and-error method. Since the neutrinosphere is a function of
both temperature and density, which oscillates with a period $\sim
0.32$ ms; therefore it also oscillates with the same period. The
temperatures and densities as a function of time are shown in Fig.~1
and Fig.~2 for the neutron star with 1.55 and 1.75 $M_{\odot}$
respectively.

From these two figures it appears that the temperature oscillation
and the density oscillation are almost close to 180$^{\circ}$ out of
phase. Since the entire system is oscillating, therefore every
quantity should oscillate with the same period. Also the equations
governing the evolution of temperature and density are not
identical, it is not surprising that they have a phase difference.
However, the question is why they are almost 180$^{\circ}$ out of
phase? This phenomenon can be explained qualitatively as follows.
When the oscillatory cycle begins, matter is falling in until the
end of the first half cycle. However, once the matter is squeezed,
the temperature is rising everywhere inside the star and the
gradient of the temperature starts to drive the thermal energy
outward. In fact most of thermal energy is generated in the core.
Furthermore the definition of the neutrinosphere is that the product
of the temperature and density is a constant. Therefore, when the
density is decreasing with time, but the temperature is increasing
with time in the first half of the cycle, there will always have a
situation that the density is minimum and the temperature is maximum
near the end of the first half of the cycle at the neutrinosphere.

Using $R_\nu$ and $T_\nu$ at the neutrinosphere obtained from the
numerical simulation, we compute the time evolution of the neutrino
luminosity. The results are shown in Fig.~3 and Fig.~4 for two
different neutron star mass respectively. We can see that extremely
intense neutrino pulses can occur when the temperature at the
neutrinosphere is maximum while the density at the neutrinosphere is
minimum. The pulsation period of the neutrino luminosity is equal to
that of the temperature and density. In calculating the neutrino
luminosity we have used the constraint $L_{\nu} < \dot{E}_M$ where
$\dot{E}_M$ represents the rate of energy flow carried by fluid into
the cell containing the neutrinosphere $R_{\nu}$. Explicitly
$\dot{E}_M = 4 \pi R_{\nu}^2 \Delta R \tau/\delta t$, where $\delta
t$ is the dynamical time scale for the energy flow from one grid to
another grid and $\tau = 0.5 \rho v^2 + \rho \epsilon$ is the total
energy density at $R_{\nu}$.

The maximum neutrino luminosities range from $10^{53}$ to $10^{54}$
erg/s. The total energies carried away by neutrinos in the period of
first 3ms are $6.75\times 10^{49}$ ergs and $7.12\times 10^{49}$ergs
for the star with 1.55M$_{\odot}$ and  for the star with
1.75M$_{\odot}$ respectively and the typical neutrino energy is $>
30$MeV during the peaks.

\subsection{Damping of the oscillation}

With the grid resolution ($dx=0.28$ km) we used for the simulations,
which is limited by the computational resource, we see that
numerical damping becomes significant after about 3 ms. Recently
Adikamalov et al (2008) have used a 2D general relativistic
numerical code to study the gravitational-wave signals emitted from
the phase-induced collapse of neutron stars and their results are
very similar to our previous results (Lin et al. 2006). Since
Adikamalov et al (2008) can achieve much higher resolution in 2D,
they can perform longer timescale simulations with high accuracy.
They conclude that the timescale of hydrodynamical damping effects
(e.g., due to mass-shedding) is typically a few tens to hundreds ms.

However, there are still other physical damping effects which are
not modelled in the simulations (Lin et al. 2006; Adikamalov et al
2008). Theoretically Wang \& Lu (1984) first pointed out that the
dissipation due to nonleptonic reaction is of great importance and
the stellar pulsations of the quark stars would be strongly damped
via the following process
\begin{equation}
s + u \leftrightarrow u + d.
\end{equation}

According  to \citet{Sawyer1989} and \citet{Madsen1992}, in the
high-temperature limit, which is exactly our case, the bulk
viscosity can be obtained analytically
\begin{equation}
\zeta=\frac{\alpha T^2}{\omega^2+\beta T^4}\Big (1-\big(1-\exp
(-\beta^{1/2} T^2 \tau)\big)\frac{2
\beta^{1/2}T^2/\tau}{\omega^2+\beta T^4}\Big)
\end{equation}

\begin{equation}
\alpha = 9.39 \times 10^{22} m_s^4 \mu_d^3 \quad \textrm{g cm}^{-1}
\textrm{s}^{-1}
\end{equation}

\begin{equation}
\beta = 7.11 \times 10^{-4} \mu_d^6 (1 + m_s^2/4\mu_d^2)^2 \quad
\textrm{s}^{-2}
\end{equation}
where $m_s$ is the mass of strange quark mass in MeV, typically
ranges from 100 to 300 MeV. $\mu_d$ is the down quark chemical
potential, the typical value is around 235 MeV, assuming the nuclear
matter density is $2.8 \times 10^{14}\, \textrm{g cm}^{-3}$. $\tau$
is the perturbation period and $\omega = 2 \pi / \tau$.

For the star with relative constant density, \citet{Sawyer1989}
estimated the damping time of vibration:
\begin{equation}\label{eq:TD}
\tau_D = 30^{-1}\rho R^2 \zeta^{-1}
\end{equation}

The average damping time is thus calculated with average temperature
($\langle T \rangle$) and density ($\langle \rho \rangle$), which are given by.
\begin{equation}
\langle T \rangle = \frac{\int^{R_{tr}}_0 4 \pi r^2 T(r) \rho(r)
dr}{\int^{R_{tr}}_0 4 \pi r^2 \rho(r) dr},
\end{equation}
\begin{equation}
\langle \rho \rangle = \frac{\int^{R_{tr}}_0 4 \pi r^2 \rho(r)
dr}{\int^{R_{tr}}_0 4 \pi r^2 dr},
\end{equation}
where $R_{tr}$ is the radius where the matter transits to quark
matter, i.e. $\rho(R_{tr})=\rho_{tr}$.

If we take $m_s \sim 140$ MeV, $\langle \rho \rangle \sim
10^{15}~{\rm g/cm}^3$ and $\langle T \rangle \sim 50$ MeV, the
damping time scale is $\sim 10$ s. This time scale is sufficient
long enough to allow the neutrinos to carry away most of
gravitational energy released from the phase-induced collapse
neutron star ($\Delta E_G \sim GM^2 \Delta R/R^2$, where $\Delta R$
is the change of radius  before and after the phase transition.).
For the model presented in this paper, $\Delta R/R$ is $\sim 0.2$,
which gives $\sim 10^{53}$ ergs, which does not conflict with the
IMB results. Furthermore, in our simulation neutrinos can carry away
$\sim 10^{50}$ ergs in 3 ms. If we assume that this is the only
mechanism to damp out the oscillation energy, it only takes $3\ {\rm
ms} \times \frac{10^{53} \ {\rm erg} }{10^{50}\ {\rm erg} } \sim 3$
s to take this amount of energy away by neutrinos with energy $> 30$
MeV.

\section{Cooling of quark stars and neutron stars}

The use of the thermally excited
helical vortex waves that produce fast magnetosonic waves in the
stellar crust, and which propagate toward the surface and transform
into outgoing electromagnetic radiation, has allowed the direct
determination of the core temperature of neutron stars
\citep{Svi03}. The core temperature of the Vela pulsar is $T=8
\times 10^8$ K, while the core temperature of PSR B0656+14 and
Geminga exceeds $2\times 10^8$ K respectively. These temperature
estimates rule out the equations of state incorporating Bose
condensations of pions or kaons and superfluid quark matter in these
objects. Thermal X-ray radiation from neutron star soft X-ray
transients in quiescence provides the strongest constraints on the
cooling rates of neutron stars, and thus on the interior composition
and properties of matter in the cores of compact objects
\citep{Hei06}. The analysis of the new (2006)and archival (2001)
XMM-Newton observations of the accreting millisecond pulsar SAX
J1808.4-3658 in quiescence provides the most stringent constraints
to date. Simultaneous fitting of all available XMM data allows a
constraint on the quiescent neutron star (0.01-10 keV) luminosity of
$L_{NS}<1.1\times 10^{31}$ erg/s. This limit excludes some current
models of neutrino emission mediated by pion condensates
\citep{Hei06}, and provides further evidence for additional cooling
processes, such as neutrino emission via direct Urca processes,
involving nucleons in the cores of massive neutron stars. Hence, these recent observations show that the observed thermal luminosity of some neutron stars does not agree with some proposed exotic cooling processes \citep{Ts98, YH03,YP04}.

\subsection{Energy dissipation mechanisms for strange stars}

Several physical processes that contribute to the energy emission
from the bare quark star surface have been proposed. The most
important of these energy loss mechanisms is the electron-positron
pair creation, due to the strong electric field at the surface of
the strange star \citep{Us98a,Us98}. Others processes include neutral pion
emission \citep{St02}, quark-quark bremsstrahlung
\citep{Chmaj,Ch03}, the equilibrium black body radiation
\citep{Chmaj} and the electron-electron bremsstrahlung \citep{Ja04, Ha05}, respectively.

In the case of the energy loss via the production of $e^{-}e^{+}$
pairs, the energy flux is given by $F_{\pm }=\varepsilon _{\pm
}\dot{n_{\pm }}$, with $\varepsilon _{\pm }=m_{e}+T$ and
\begin{equation}\label{usov}
\dot{n_{\pm}}=\Delta r\frac{ 9T^3}{2\pi \varepsilon _{F}^{2}}
\sqrt{\frac{\alpha }{\pi }}e^ { -2m_{e}/T} n_{e}J\left( \xi
\right),
\end{equation}
where $\varepsilon _{F}=\left( \pi ^{2}n_{e}\right) ^{1/3}$ is the
Fermi momentum of the electrons with number density $n_e$, $\alpha $ is the fine 
structure constant, $\xi =2\sqrt{\alpha /\pi }\left( \varepsilon _{F}/T\right) $,
$J\left( \xi \right) =\left( 1/3\right) \xi ^{3}\ln \left( 1+2\xi
^{-1}\right) /\left( 1+0.074\xi \right) ^{3}+\left( \pi
^{5}/6\right) \xi ^{4}/\left( 13.9+\xi \right) ^{4}$, and $\Delta
r$ is the thickness of the electron layer \citep{Us01}.
\citet{Ha06} have considered the mean value of the
$z$-dependent chemical potential, $\varepsilon _{F}\approx
\left\langle \mu _{e}\left(
z,T\right) \right\rangle \approx \left( 1/d\right) \sqrt{3/\alpha \pi }%
\left( V_{I}/T\right) \left( 1+\sqrt{1+V_{I}^{2}/2\pi
^{2}T^{2}}\right) ^{-1} $. For temperatures of around
$T=10$ MeV we obtain $\varepsilon _{F}\approx 3 $ MeV, a smaller
value than the one considered in \citet{Us01}, $\varepsilon
_{F}\approx 18$ MeV. This choice of the Fermi momentum
significantly reduces the energy flux from the electron-positron
pair creation.

Pions are created at the stellar surface due to the collision
between the quarks and the bag, representing the conversion of the
quark kinetic energy to the mass of the
pion cloud \citep{St02}. The pions are assumed to decay via the processes $%
\pi ^{0}\rightarrow 2\gamma \longleftrightarrow e^{+}+e^{-}$ and
$\pi ^{\pm }\rightarrow \mu ^{\pm }+\nu _{\mu }\rightarrow e^{\pm
}+\nu _{e}+2\nu _{\mu }$. The energy flux due to this process is
given by $F_{\pi }=\rho _{\pi }v_{\pi }$, where $\rho _{\pi }$ is
the energy density of the pion field at
the star's surface, which is fixed by the axial current conservation, and $%
v_{\pi }$, the speed of the emitted pions, is $v_{\pi }\approx \sqrt{%
2T/m_{\pi }}$, where $m_{\pi }\approx 140$ MeV is the mass of the
pion \citep{St02}.

Electron-electron bremsstrahlung is an important energy loss
mechanism for strange stars \citep{Ja04,Ha05}. For surface
temperatures $T<10^9$ K, the photon flux exceeds that of the
electron-positron pairs that are produced via the Schwinger
mechanism in the presence of a strong electric field that binds
electrons to the surface of the quark star. The average energy of
photons emitted from the bremsstrahlung process can be 0.5 MeV or
more, which is larger than that in electron-positron pair
annihilation. The effect of the multiple and uncorrelated
scattering on the radiation spectrum (the
Landau-Pomeranchuk-Migdal effect), together with the effect of the
strong electric field at the surface of the star, was discussed in
\citet{Ha05}. The presence of the electric field strongly
influences the radiation spectrum emitted by the electrosphere.
The radiation properties of the electrons in the electrosphere
essentially depend on the value of the electric potential at the
quark star surface. The effect of the multiple scattering, which
strongly suppresses radiation emission, is important only for the
dense layer of the electrosphere situated near the star's surface,
and only for high values of the surface electric potential of the
star. The bremsstrahlung emissivity of the electrosphere can be
obtained as
\begin{equation}\label{emis}
\varepsilon _{Br}^{(ee)}\left(\mu ,T\right) =\frac{75}{4\pi
^{2}}\alpha
r_{e}^{2}g^{2}{\rm Li}_{7/2}\left( \eta \right) \left[ \left( \ln \frac{T}{2m_{e}}+
\frac{77}{30}-\gamma \right) {\rm Li}_{7/2}\left( \eta \right) +
\frac{d{\rm Li}_{n}\left( \eta \right) }{dn}\bigg{|} _{
n=7/2}\right]T^{7},
\end{equation}
where $g=2$ is the statistical weight for electrons,
$\eta =-\exp\left[ \mu _{e}\left(T\right) /T\right]$,
${\rm Li}_{n}\left(\eta \right)=\sum_{k=1}^{\infty }\eta ^{k}/k^{n}$
is the polylogarithm function and
$\gamma =0.577216$ is Euler's constant.

The bremsstrahlung energy flux from the exterior electron layer of
the strange star, coming out from a thin surface layer of thickness $dz$, is
$F_{Br}^{(ee)}=\varepsilon _{Br}^{(ee)}dz/\pi $. Taking into
account the contribution of all layers we find
\begin{equation}
F_{Br}^{(ee)}(T)=\frac{1}{\pi }\int_{0}^{\infty }\varepsilon
_{Br}^{(ee)}\left( z,T\right) dz=\sigma _{Br}^{(ee)}(T)T^{7},
\end{equation}
where
\begin{eqnarray}
\sigma _{Br}^{(ee)}(T)&=&\frac{75}{4\pi ^{3}}\alpha
r_{e}^{2}g^{2}\times \nonumber\\
&&\int_{0}^{\infty }{\rm Li}_{7/2}\left[ \eta \left( z,T\right)
\right]
\left\{ \left( \ln \frac{T}{2m_{e}}+\frac{77}{30}-\gamma \right) {\rm Li}_{7/2}
\left[ \eta \left( z,T\right) \right] +\frac{d{\rm Li}_{n}\left(
\eta \right) }{dn}\bigg{|} _{ n=7/2}\right\} dz.\nonumber\\
\end{eqnarray}

The energy flux from the quark-quark bremsstrahlung can be represented as $%
F_{q-q}=\sigma _{Br}\left( n_{b},T\right) T^{4}$, where $\sigma
_{Br}\left( n_{b},T\right) \approx g^{2}Li_{2}^{2}\left[ -\exp
\left( \left( \pi ^{2}n_{b}\right) ^{1/3}/T\right) \right] I\left(
n_{b},T\right) \lambda /\left( 2\pi \right) ^{3}$, $\lambda $ is
the mean photon path in the quark matter, $n_{b}$ is the baryon
number density at the surface of the star and
\begin{eqnarray}\label{quarkquark}
\frac{\pi }{\alpha }I\left( n_{b},T\right)  &\approx &\frac{1+3\pi
/4\alpha
_{s}}{\tau }\ln \left[ \frac{1+4\left( 1+3\pi /4\alpha _{s}\right) ^{2}}{%
\left( 1+4\tau ^{2}a^{2}n_{b}^{2/3}\right) ^{an_{b}^{1/3}}}\right]
+\nonumber\\
&&\frac{1}{\tau }\left[ \arctan 2\left( 1+3\pi /4\alpha
_{s}\right) -\arctan
2\tau an_{b}^{1/3}\right] -  \nonumber \\
&&4\left( \frac{1+3\pi /4\alpha _{s}}{\tau }-an_{b}^{1/3}\right) +\frac{1}{%
2\tau }\left[ D\left( 1+3\pi /4\alpha _{s}\right) -D\left( \tau
an_{b}^{1/3}\right) \right],
\end{eqnarray}
where $\alpha _{s}$ is the strong coupling constant,
$a=2^{1/3}3^{4/3}\pi
^{5/3}g^{2/3}e^{4}Z^{2}\ln \left( 184Z^{-1/3}\right) /\alpha _{s}^{2}$, $%
\tau $ is the mean collision time between quarks, given by
\begin{equation}
\tau
^{-1}\approx -n_{b}^{-1/3}T^{2}{\rm Li}_{2}\left[ -\exp \left(
\left( \pi ^{2}n_{b}\right) ^{1/3}/T\right) \right] /8\pi ^{8/3},
\end{equation}
which can be roughly
estimated as $\tau =1/n\sigma _{0}v$, and the function $D(x)$ is defined as $%
D(x)=i\left[ {\rm Li}_{2}\left( -2ix\right) -{\rm Li}_{2}\left(
2ix\right) \right] $ \citep{Ch03}. $\sigma _{0}$ is the
quark-quark elastic scattering cross section. Due to interference
between amplitudes of nearby interactions, the bremsstrahlung
emissivity from strange star surface is suppressed for frequencies
smaller than a critical frequency (the Landau-Pomeranchuk-Migdal
effect) \citep{Ch03}. The range of the suppressed frequencies is a
function of the quark matter density at the star's surface and of
the QCD coupling constant. For temperatures much smaller than the
Fermi energy of the quarks the bremsstrahlung spectrum has the
same temperature dependence as the equilibrium black body
radiation. Multiple collisions could reduce the intensity of the
bremsstrahlung radiation by an order of magnitude. The absorption
in the semi-degenerate electron gas can also significantly reduce
the intensity of the quark-quark bremsstrahlung radiation and,
consequently, the surface emissivity. The combined effects of
multiple collisions and absorption in the electron layer could
make the soft photon surface radiation of quark stars six orders
of magnitude smaller than the equilibrium black body radiation
\citep{Ch03}.

At temperature $T$ strange matter is filled with electromagnetic
waves in thermodynamic equilibrium with quarks. The quanta of the
electromagnetic waves in plasma (transverse plasmons) have a
characteristic dispersion relation $\omega \left( k\right)
=\sqrt{\omega _{p}^{2}+k^{2}}$, where $k$ is the wave-number and
the plasma frequency $\omega _{p}=e\left( 8\pi n_{b}/3\mu \right)
^{1/2}$ \citep{Chmaj}. The characteristic transverse plasmon
cutoff frequency can be estimated as $\omega _{p}\approx 20-25$
MeV \citep{Chmaj,St02}. A beam of transverse plasmons hitting the
edge of strange star matter from inside will be partially
reflected and partially will pass to the outer vacuum, being
refracted. The energy flux of thermal
equilibrium photons radiated from the bare strange surface is given by
\begin{equation}\label{equil}
F_{eq}=\frac{1}{2} \int_{\omega _{p}}^{\infty }\omega \left(
\omega ^{2}-\omega _{p}^{2}\right) g\left( \omega \right) \left[e^
{\omega /T} -1\right] ^{-1}d\omega ,
\end{equation}
where $g\left( \omega \right) =\left( 1/2\pi ^{2}\right)
\int_{0}^{\pi /2}\left[ 1-\left( R_{\perp }+R_{\parallel }\right)
/2\right] \sin \theta \cos \theta d\theta $, with $R_{\perp }=\sin
^{2}\left( \theta -\theta _{0}\right) /\sin ^{2}\left( \theta
+\theta _{0}\right) $ and $R_{\parallel }=\tan ^{2}\left( \theta
-\theta _{0}\right)
/\tan ^{2}\left( \theta +\theta _{0}\right) $. $\theta _{0}$ is defined as $%
\theta _{0}=\arcsin \left[ \sin \theta \sqrt{1-\left( \omega
_{p}/\omega \right) ^{2}}\right]$, respectively \citep{Chmaj}.

The variation of the energy fluxes $F$ for these energy mechanisms
are represented, as a function of the temperature, in Fig.~5.

When the temperature of the quark star core drops below $10^{9}$ K,
the strange matter becomes superfluid. At this temperature quarks
can form colored Cooper pairs near the Fermi surface and become
superconducting. From the BCS theory it follows that the critical
temperature $T_{c}$ at which the transition to the superconducting
state takes place is $T_{c}=\Delta /1.76$, where $\Delta $ is the
pairing gap energy \citep{Bl}. An early estimation of $\Delta $ gave
$\Delta \sim 0.1-1$ MeV \citep{early}, but some recent studies
considering instanton-induced interactions between quarks estimated
$\Delta \sim 100$ MeV \citep{all1a,all1b}. When the temperature of
the star is below $T_{c}$, the emissivity is of the quark matter is
modified by the superconducting effects. Since the collisions
between the quarks and the bag are suppressed, pion emissivity is
suppressed by a factor of $\exp \left( -\Delta /T\right) $
\citep{St02}. The same factor can be used to describe the
suppression of the quark-quark bremsstrahlung due to the
superconductivity of the strange matter.

Fig.~6 shows the variation with temperature of the energy fluxes for
various dissipation mechanisms if quarks become superfluid.

For $T<<\omega_p$ the equilibrium photon emissivity of strange
matter is negligible small, compared to the black body spectrum.
Even for a small value of the gap pairing energy $\Delta $ the
superfluidity of the quark matter strongly suppresses the
quark-quark bremsstrahlung and pion emissivity of the star.

Strange quark matter in the color-flavor locked (CFL) phase of
QCD, which occurs for large gaps ($\Delta \sim 100$ MeV), could
be rigorously electrically neutral, despite the unequal quark
masses, even in the presence of the electron chemical potential
\citep{all1a,all1b}.

However, \citet{PaUs02} pointed out that for sufficiently large
$m_{s}$ the low density regime is rather expected to be in the
2-color-flavor Superconductor'' phase in which only $u$ and $d$
quarks of two color are paired in single condensate while the ones
of the third color and $s$ quarks of all three colors are
unpaired. In this phase, electrons are present. In other words,
electrons may be absent in the core of strange stars but present,
at least, near the surface where the density is lowest.
Nevertheless, the presence of CFL effect can reduce the electron
density at the surface and hence also significantly reduces the
bremsstrahlung emissivity of the electrons in the surface layer.

\subsection{Cooling of strange stars}

In this subsection, we consider the cooling of a quark star. For simplicity, we assume that the star is of uniform
density and isothermal, that is, the core and surface temperatures
are equal. The effect of the magnetic field on the cooling is also
neglected. The thermal evolution of the quark star is determined
by the equation
\begin{equation}
C_{V}\frac{dT}{dt}=-\sum_{i=1}^{n}L_{i}=-\left(
L_{pair}+L_{qq}+L_{ee}+L_{\pi}+L_{eq}+L_{\nu }\right) ,
\end{equation}
where $C_{V}$ is the specific heat, $T$ is the temperature, and
$L_{pair}$ is the electron-positron pair luminosity, $L_{qq}$ is the
quark-quark bremsstrahlung luminosity, $L_{ee}$ is the
electron-electron bremsstrahlung luminosity, $L_{\pi}$ is the pion
emission luminosity, $L_{eq}$ is the luminosity due to the thermal
equilibrium radiation, and $L_{\nu }$ is the neutrino luminosity,
respectively. In the normal state of quark matter, the quark Fermi
momentum $p_{Fq}$ can be approximated as $p_{Fq}=235\left( \rho
/\rho _{0}\right) ^{1/3}$
MeV/$c$, where $\rho _0$ is the nuclear density, and the specific heat of the quark 
matter is given by $%
c_{q}=2.5\times 10^{20}\left( \rho /\rho _{0}\right) ^{2/3}T_{9}$ erg cm$%
^{-3}$ K$^{-1}$ \citep{Iw80}. In the superfluid state, for
$0.2T_{c}\leq T\leq T_{c}$, the specific heat can be obtained as
$c_{q}^{(sf)}=3.15c_{q}\left( T_{c}/T\right) \exp \left(
-1.76T_{c}/T\right) \left[ 2.5-1.66\left(
T/T_{c}\right) +3.64\left( T/T_{c}\right) ^{2}\right] $, while for $%
T<0.2T_{c}$ the specific heat is zero (see \citet{St02} and
references therein). The Goldstone excitations in the quark-gluon
plasma contribute to the specific heat of the strange star, so that
$c_{g-\gamma }=3.0\times 10^{13}N_{g-\gamma }T_{9}^{3}$ erg
cm$^{-3}$ K$ ^{-1}$, where $N_{g-\gamma }$ is the number of
available massless gluon-photon states, which are present even in
the color superconducting phase \citep{Bl}. When the temperature is
low, the specific heat of quark matter vanishes. However, the
electrons are not affected, and at low temperatures their effect is
important. The specific heat of the electrons is given by $
c_{e}=1.7\times 10^{20}\left( Y_{e}\rho /\rho _{0}\right)
^{2/3}T_{9}$ erg cm $^{-3}$ K$^{-1}$ \citep{St02}, where $Y_e
\approx 10^{-3}$ is the electron fraction.

Neutrinos are emitted by quark matter through the URCA process,
$d\rightarrow u+e^{-}+\bar{\nu} _{e^{-}}$, $u+e^{-}\rightarrow d+\nu
_{e^{-}}$. The neutrino emissivity for this process can be obtained
as $\varepsilon _{d}\simeq 8.8\times 10^{26}\alpha _{c}\left( \rho
/\rho _{0}\right) Y_{e}^{1/3}T_{9}^{6}$ erg cm$^{-3}$ s$^{-1}$
\citep{Iw80}. In the superfluid state the neutrino emissivity is
suppressed by a factor of $\exp \left( -\Delta /T\right) $. The
exact form for the neutrino emissivity can be obtained as
$\varepsilon _{d}=\left( 457\pi /840\hbar ^{10}c^{9}\right)
G_{F}^{2}\cos ^{2}\theta _{C}\left( 1-\cos \theta _{ue}/a\right)
P_{Fu}P_{Fd}P_{Fe}\left( k_{B}T\right) ^{6}$ erg cm$^{-3}$ s$^{-1}$
\citep{Du83}, where $G_{F}$ is the Fermi weak coupling constant,
$\theta _{C}$ is the Cabibbo angle, $a=\left( 1-2\alpha _{c}/\pi
\right) ^{-1/3}$, $P_{Fu}$, $P_{Fd}$ and $P_{Fe}$ are the Fermi
momenta of the $u$, $d$ quarks, and of the electrons, respectively.
The angle between the $u$ and $e$ momenta, $\theta _{ue}$ is given
by $\cos \theta _{ue}=\left( P_{Fd}^{2}-P_{Fu}^{2}-P_{Fe}^{2}\right)
/2P_{Fu}P_{Fe}$. The condition for the URCA process to occur is
$\left| \cos \theta _{ue}\right| \leq 1$ \citep{St02}. In the case
of the similar reactions for the $s$ quarks,  $s\rightarrow
u+e^{-}+\bar{\nu}_{e^{-}}$, $u+e^{-}\rightarrow s+\nu _{e^{-}}$, the
condition $\left| \cos \theta _{ue}^{^{\prime }}\right| \leq 1 $,
where the angle $\theta _{ue}^{^{\prime }}$ is defined as $\cos
\theta _{ue}^{^{\prime }}=\left(
P_{Fs}^{2}-P_{Fu}^{2}-P_{Fe}^{2}\right) /2P_{Fu}P_{Fe}$, is
satisfied only if the density of the quark matter is as high as
$4\times 10^{16}$ g cm$^{-3}$. Therefore these reactions do not
exist in the quark stars. The neutrino emissivity due to the photon
gluon mixing, which can generated a massive photon-gluon excitation,
is neglected, since it is important only at temperatures higher than
$70$ MeV \citep{Bl}. The cooling curve and the luminosity curve of
the 1.55M$_{\odot}$ strange star are presented as the solid curve in
Fig.~7 and Fig.~8 respectively. We can see the radiation luminosity
at $t\sim 20$ years is below $10^{34}erg/s$ but with a high
temperature $T \sim 10^7$K and the spectrum is a thermal
bremsstrahlung.

\subsection{Neutron star cooling process}

In this subsection we describe a simple cooling model of neutron
stars including possible exotic matter core, e.g. meson condensate
or quark core (see e.g. Cheng et al. 1992, Chong \& Cheng 1993). We
assume that the neutron star with 1.55M$_{\odot}$ and $10^6$ cm in
radius is uniform density and has a isothermal core temperature
($T_c$). We adopt the simple relation between the core temperature
and the surface temperature ($T_s$) given by the relation $T_{c
8}=1.3 (T_{s 6}^4 / g_{s 14})^{0.445}$ \citep{Gud81}. The core
temperature evolution of the star is given by
\begin{equation}
C_v dT_c/dt = -L_{\nu} - L_{bb}(T_s) \textrm{ .}
\end{equation}
The heat capacity $C_v$ is given by \citep{Max79}
\begin{equation}
C_v = C^{(e)} + C^{(n)}
\end{equation}
where
\begin{equation}
C^{(e)} = 1.9 \times 10^{37} M_\odot \rho_{14}^{1/3} T_9 \textrm{ erg K}^{-1}
\end{equation}
is the electron heat capacity with $\rho_{14} = \rho / 10^{14}$ g
cm$^{-3}$ and $T_9 = T/10^{9}$ K, and
\begin{equation}
C^{(n)} = \left\{ \begin{array}{l l}
 C_n(T) = 2.3 \times 10^{39} M_\odot \left( \frac{m_n^*}{m_n}\right)  \rho_{14}^{-2/3}
T_9 \textrm{ erg K}^{-1} & \textrm{ for } T> T_n \\
3.15 C_n(T) \left( \frac{T_n}{T}\right)  e^{-1.76 T_n / T} \left[
2.5 - 1.66 \frac{T}{T_n} + 3.64 \left( \frac{T}{T_n}\right) ^2
\right] \textrm{ erg K}^{-1} &
\textrm{ for } T \le T_n \\
\end{array} \right. \\
\end{equation}
is the neutron heat capacity, with $T_n$ the transition temperature
of normal-superfluid neutrons and $m_n^*$ the effective mass of
neutrons.  We adopted $m_n^*/m_n \simeq 1$ and $T_n = 3.2 \times
10^9$ K \citep{TaTa71}.

The major neutrino emissions are due to modified Urca process and
NN-bremsstrahlung, where N is a ncleon, n or p.  These mechanisms
are relatively weak and the luminosity can be approximated by
(Yakovlev \& Haensel 2003)
\begin{equation}
L_\nu = L_s T_9^8
\end{equation}
where $L_s$ is at the range of $4 \times 10^{38}$ to $10^{40}$ erg s$^{-1}$.

When considering some exotic neutrino processes, the emission is
greatly enhanced.  The neutrino luminosity can be written
as \citep{YH03}
\begin{equation}
L_\nu = L_f T_9^6
\end{equation}
where $L_f$ lies at the range $4 \times 10^{41}$ to $4 \times
10^{44}$  erg s$^{-1}$ for pion condensation process and $4 \times
10^{41}$ to $4 \times 10^{43}$  erg s$^{-1}$ for kaon condensation
and quark cooling process.

The cooling curves and the radiation luminosities of neutron star
with various cooling mechanisms are shown in Fig.~7 and Fig.~8
respectively. We can see that the neutron star without exotic
cooling processes, e.g. pion condensation or kaon condensation,
their thermal luminosity is much larger than $10^{34}$ erg/s. But the
the exotic cooling processes can cool the star much faster and the
thermal luminosity can be much lower than $10^{34}$erg/s. On the
other hand, the surface temperature of the neutron star with exotic
cooling processes is very low $< 10^6$K and it is thermal spectrum.

\section{Discussions and final remarks}

The detection of neutrinos from SN1987A has suggested that a compact
object should be formed at the center of SN1987A. If the time delay
and energy distribution difference between neutrinos detected by
KAM-II and IMB are genuine together with the absence of hot compact
object at the center of 1987A,which implies a rapidly cooling cold
compact,  we have proposed that the neutrinos detected by IMB may
result from a delay phase-induced collapse after the neutron star is
formed. The final compact object may be a neutron star with exotic
matter core or a strange star, either of them is a rapidly cooling
object. Consequently the current radiation luminosity from SN1987A
should be much lower than $10^{34}$ erg/s. However, we have argued
that the phase transition from neutron star to strange star may be
the most possible phase transition occurred in 1987A. Such collapse
process can generate temperature and density oscillation on the
neutrinosphere. By using a three dimensional Newtonian code with a
toy model, in which the phase transition is taken to be
instantaneous and the equation of state is assumed to be a mixed
phase, we find that intense pulsating neutrinos with energy $>$30MeV
can be emitted. This phenomenon should be generic because the
compression generates more heat in the core and the heat flow is
outgoing during the contraction phase, whereas the matter is flowing
in. That should create a situation that when the temperature is
maximum and the density is minimum at the neutrinosphere. It is very
interesting to detect intense pulsating hot neutrino emission from
supernova explosions occurring from nearby galaxies. The future
neutrinos experiments, like ANTARES, and IceCube (see \citet{Hal06}
for a review) may be able to detect such signals.

In calculating the cooling of strange star, we have assumed a bare
strange star, i.e. a strange star without crust. This assumption may
be justified by the following reasons: On the top of the quark
matter, a thin crust can exist as long as the electron density
within it is smaller than that in the quark matter \citep{AlFa86}.
Such a baryonic crust is, however, much thinner than the neutron
star crust. The mass of the crust on the quark star is at most $
10^{-5}M_{\odot }$, because the density of the crust bottom cannot
exceed neutron drip density, and it may extend to up to $250-300$ m.
The normal crust may occur, for instance, due to the accretion of
normal matter onto a bare strange star. The oscillation spectrum of
strange stars with crust differs from the spectrum of neutron stars
\citep{Chug07}. If detected, acoustic oscillations would allow one
to discriminate between strange stars with crust and neutron stars,
and constrain the mass and radius of the star. The recent detection
of seismic vibrations in the aftermath of giant flares from two
magnetars (highly magnetized compact stars) is a major breakthrough
in observational astronomy. The oscillations excited seem likely to
involve the stellar crust, the properties of which differ
dramatically for strange stars. The resulting mode frequencies for
strange stars cannot be reconciled with the observations, for
reasonable magnetar parameters \citep{WaRe07}. On the other hand,
young and hot strange stars are the source of a powerful pair wind,
consisting of electron-positron pairs and photons, created by the
Coulomb barrier at the quark surface \citep{AkMiUs05}.  Photons
dominate in the emerging emission, and the emerging photon spectrum
is rather hard and differs substantially from the thermal spectrum
expected from a neutron star with the same luminosity. The total
luminosity in the case of a stationary, spherically outflowing, pair
winds is in the range of $L=10^{35}-10^{42}$ ergs/s. These results
have direct relevance to the emission from hot, bare, strange stars.
For $L>2\times 10^{35}$ erg/s, photons dominate the emerging
emission. As $L$ increases from $10^{35}$ to $10^{42}$ ergs/s, the
mean photon energy decreases from $\sim 400-500$ keV to $40$ keV,
while the spectrum changes in shape from a wide annihilation line to
being nearly blackbody with a high energy ($>100$ keV) tail. Such a
correlation of the photon spectrum with the luminosity, together
with the fact that super-Eddington luminosities can be achieved,
might be a good observational signature of hot, bare, strange stars
\citep{AkMiUs03}. At the moment of its formation, a strange star is
very hot. The temperature of the interior may be as high as a few of
$10^{11}$ K. The neutrino luminosity of the young quark star is of
the order of $10^{54}$ erg/s, the rate of mass ejection from such a
hot compact object is very high, and the normal matter envelope is
blown away by radiation pressure in a few seconds. High temperatures
also lead to a considerable reduction of the Coulomb barrier,
increasing the tunneling of nuclei through the barrier toward the
surface. Therefore, it is natural to expect that the surface of a
young strange star to be nearly (or completely) bare \citep{Us01}.
Since the strange quark matter at the surface of the star is bound
via strong interactions rather than gravity, such a star can radiate
at the at the luminosity greatly exceeding the Eddington limit.
Therefore, due to the radiation pressure, a normal matter crust
cannot be built around a young strange star.

We have also assumed that the phase transition should take place
after the neutron star is formed. We argue that the delay time scale
of the phase transition is of order of the neutrino trapping time
scale based on the following reasons: The theory of type II
supernovae predicts the gravitational collapse of a degenerate core
on timescales of the order of $t_{coll}\sim \left( G\rho \right)
^{-1/2}\sim 4\times 10^{-3}\rho _{12}s\sim 0.1$ s, followed by the
formation of a hot ($k_{B}T\sim 50$ MeV) protoneutron star, with a
radius of around $\ 50$ km, composed of $N_{b}\sim 10^{57}$ baryons,
and with the electron lepton number $L_{e}\approx 0.35N_{b}$
\citep{Ha07}. The initial mass of the collapsing core is
$M_{core}\approx N_{b}m_{0}$, where $m_{0}=1.6586\times 10^{-24}$ g
is the mass of the $^{56}$Fe nucleus, divided by 56. The
gravitational mass of the protoneutron star is only slightly lower
than $M_{core}$, because the gain in the gravitational energy is
compensated by the increase of the internal energy contained mostly
in the strongly degenerate neutrinos $\nu _{e}$, trapped in the
stellar interior. The diffusion time-scale of the neutrinos may be
estimated by assuming that coherent scattering is the dominant
opacity source, so that $t_{diff}\approx \lambda
_{A}^{coh}N_{scatt}/c$, where $\lambda _{A}^{coh}$ is the mean free
path in the sea of heavy nuclei ($A,Z$) and $N_{scatt}>1$ is the
number of scatterings experienced by the neutrino prior to escape.
By taking into account the explicit expressions of the neutrino mean
free path and of the number of scatterings, the neutrino diffusion
time can be written as $ t_{diff}\sim 0.08\rho _{12}$ s. At
sufficiently high densities $ t_{diff}>>t_{coll}$ \citep{ShTe83}.
The two timescales becomes comparable at $t_{diff}\sim t_{coll}$, a
condition which is satisfied at a density $\rho _{trap}\sim
1.4\times 10^{11}$ g/cm$^{3}$.

Neutrino trapping has enormous implications for the core collapse.
For $\rho \geq \rho _{trap}$, most of the neutrinos from electron
capture remain in the matter, and the lepton number per baryon does
not change. The neutrino distributions approaches an equilibrium
Fermi-Dirac distribution. The neutrino luminosities are greatly
reduced by trapping. Once the center of the core exceeds nuclear
densities of the order of $\rho _{nucl}=2.8\times 10^{14}$
g/cm$^{3}$, thermal pressure and nuclear forces cause the equation
of state of the matter to stiffen, preventing further collapse. Most
of the gravitational binding energy of the core is released in the
form of neutrinos, following the collapse to nuclear densities. In
the absence of neutrino trapping, the total binding energy would be
completely emitted as neutrinos in a collapse timescale, the time
for the core to contract from $ 2R_{nucl}$ to $R_{nucl}$, where
$R_{nucl}\sim 12$ km for $M\sim M_{\odot}$. Accordingly, the
neutrino luminosity would then achieve its maximum possible value,
$L_{\nu ,\max }=GM^{2}/R_{nucl}/t_{coll}\sim 10^{57}$ erg/s
\citep{ShTe83}. In reality, neutrino trapping forces the liberated
gravitational potential energy to be emitted on a much longer
diffusion timescale, $ t_{diff}>>t_{coll}$ at $\rho \sim \rho
_{nucl}$. As a result, the actual neutrino luminosity is of the
order $L_{\nu }=GM^{2}/R_{nucl}/t_{diff}\sim 10^{52}$ erg/s. Thus
during the advanced stages of the collapse, the neutrinos are unable
to stream freely out of the core. The bulk of the liberated
gravitational energy must therefore be converted into other forms of
internal energy (e.g. thermal energy, energy of the excited nuclear
states, bounce kinetic energy etc.), rather than being released
immediately in the form of escaping neutrinos.

In order for the phase transition to take place in the neutron star,
the density inside the core must satisfy the condition $\rho \geq
\rho _{trans}$ , as discussed in Section 2. Since the core
temperature during the neutrino trapping period is very high, the
thermal pressure will give a significant contribution to the total
pressure, thus lowering the central density of the neutron star, so
that $\rho <\rho _{trans}$. In order to satisfy the thermodynamical
conditions for the phase transition, the core must cool down so that
the core density increases. To estimate the density increase
resulting from the temperature decrease in the core of the neutron
star we solve the TOV equation by using a temperature dependent EOS,
i.e. Shen et al. (1998), to obtain core density as function of
temperature for a fixed stellar mass. In implementing this EOS, we
have added an ideal electron gas contribution. We have also imposed
the condition of beta equilibiurm to determine the electron
fraction. The left panel of Fig.9 shows the energy density profiles
for a star model with $M=1.55 M_\odot$ at different temperatures.
Note that the temperature profiles are determined by the general
relativistic isothermal condition. The right panel of Fig.9 shows
the results for a neutron star with $M=1.7 M_\odot$. In Fig.9, we
can see that indeed when core temperature decreases by a factor of
2, the core energy density can increase by more than 20\%. We can
see that these changes are sufficiently large to trigger the phase
transition when the core temperature decreases after the neutrino
trapping time scale.

We conclude that the abnormally high energy neutrinos and the time
delay detected by the IMB detector can be explained by the neutrino
trapping in the newly formed very hot neutron star, followed by a
phase transition. After phase transition if the compact object is a
neutron star with exotic matter core, its surface luminosity is
lower than $10^{34}$erg/s and it should have black body thermal
spectrum with temperature $<10^6$K. On the other hand if it is a
bare strange star, its surface luminosity is also lower than
$10^{34}$erg/s but the radiation spectrum is a thermal
bremsstrahlung spectrum with temperature $>10^7$K. However, the
nature of the phase transition cannot be inferred from the present
data, since the observed features cannot differentiate or
discriminate between the different phase transition models. We
suggest that the future observations on spectral features of the
compact remnant of 1987A could provide better clues to elucidate the
nature of the phase transition that might take place in SN1987A.

\acknowledgments We thank Profs. Z.G. Dai, P. Haensel, Y.F. Huang,
Kwong Lau, K.B. Luk, C. S. Pun and V. Usov for useful comments, and
the anonymous referee for very useful suggestions. This work is
partially supported by a RGC grant of the Hong Kong government of
the SAR of China under HKU7013/06P. LML acknowledges support from
the Hong Kong Research Grants Council (Grant No. 401807). The
computations were performed on the Computational Grid of the Chinese
University of Hong Kong and the High Performance Computing Cluster
of the University of Hong Kong.

\begin{figure}[!ht]
\centering \label{} \plottwo{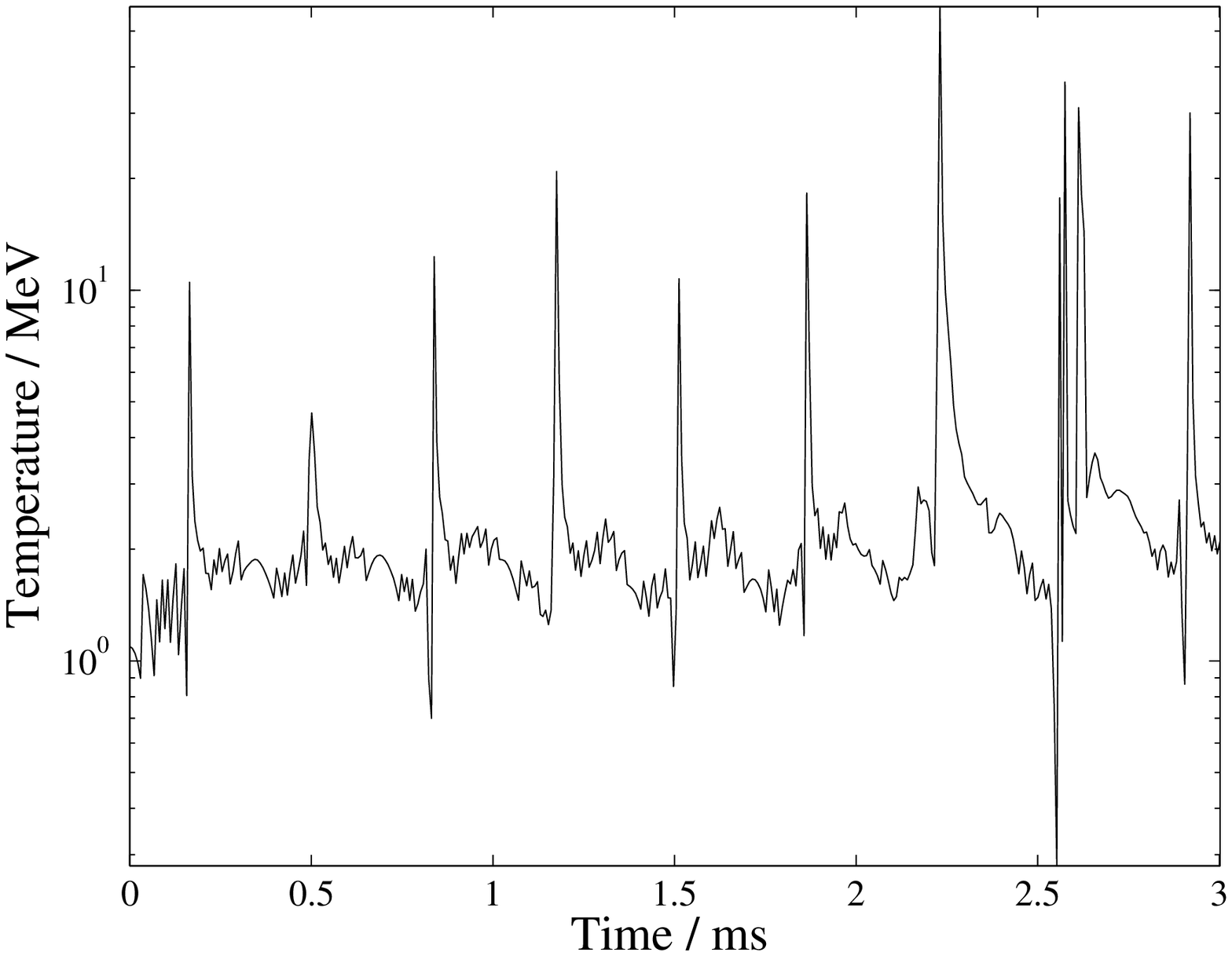}{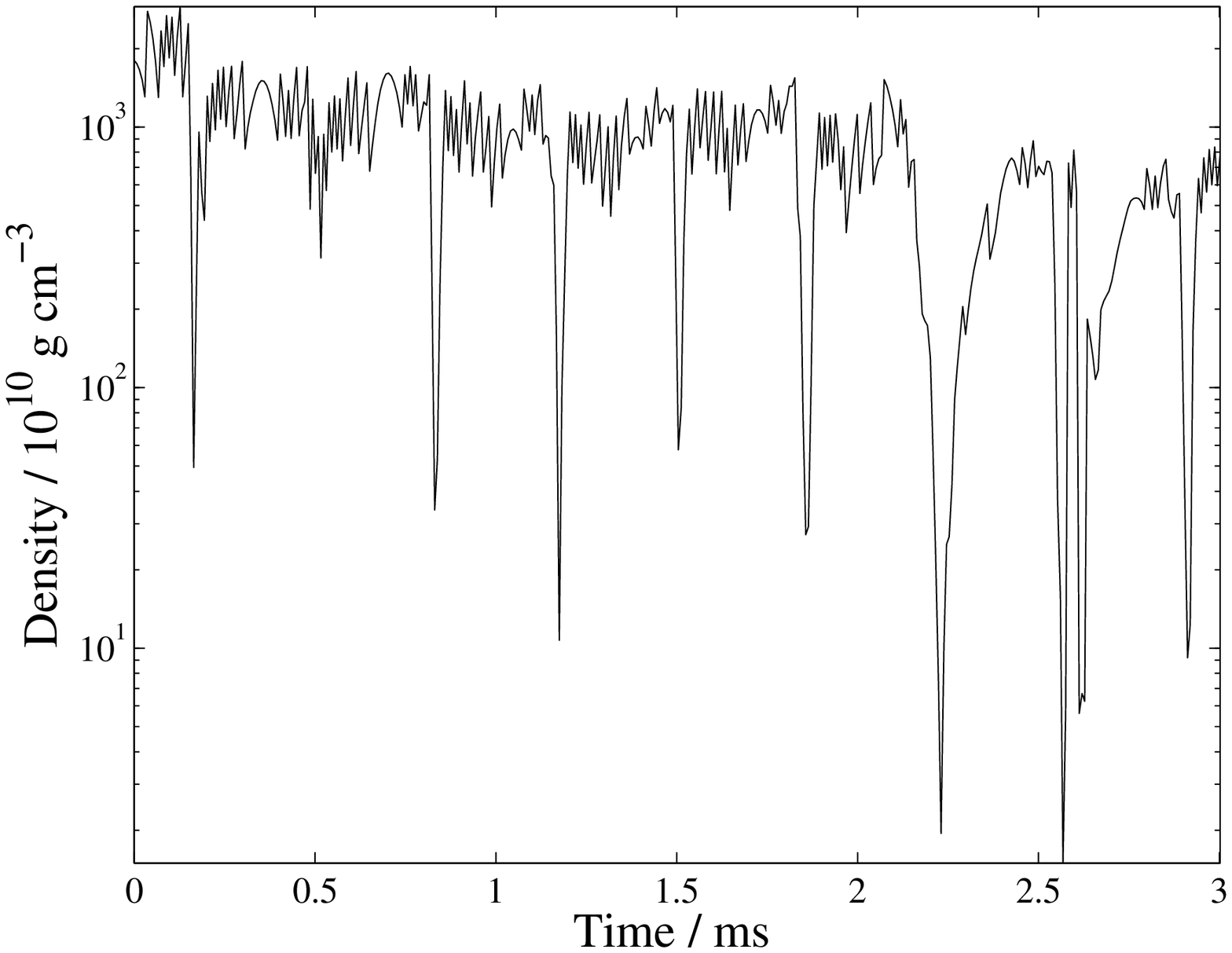}
\caption{{\itshape Left:\/} Neutrinosphere temperature versus time.
{\itshape Right:\/} Neutrinosphere density versus time. The initial
mass of neutron is 1.55M$_{\odot}$}
\end{figure}

\begin{figure}[!ht]
\centering \label{} \plottwo{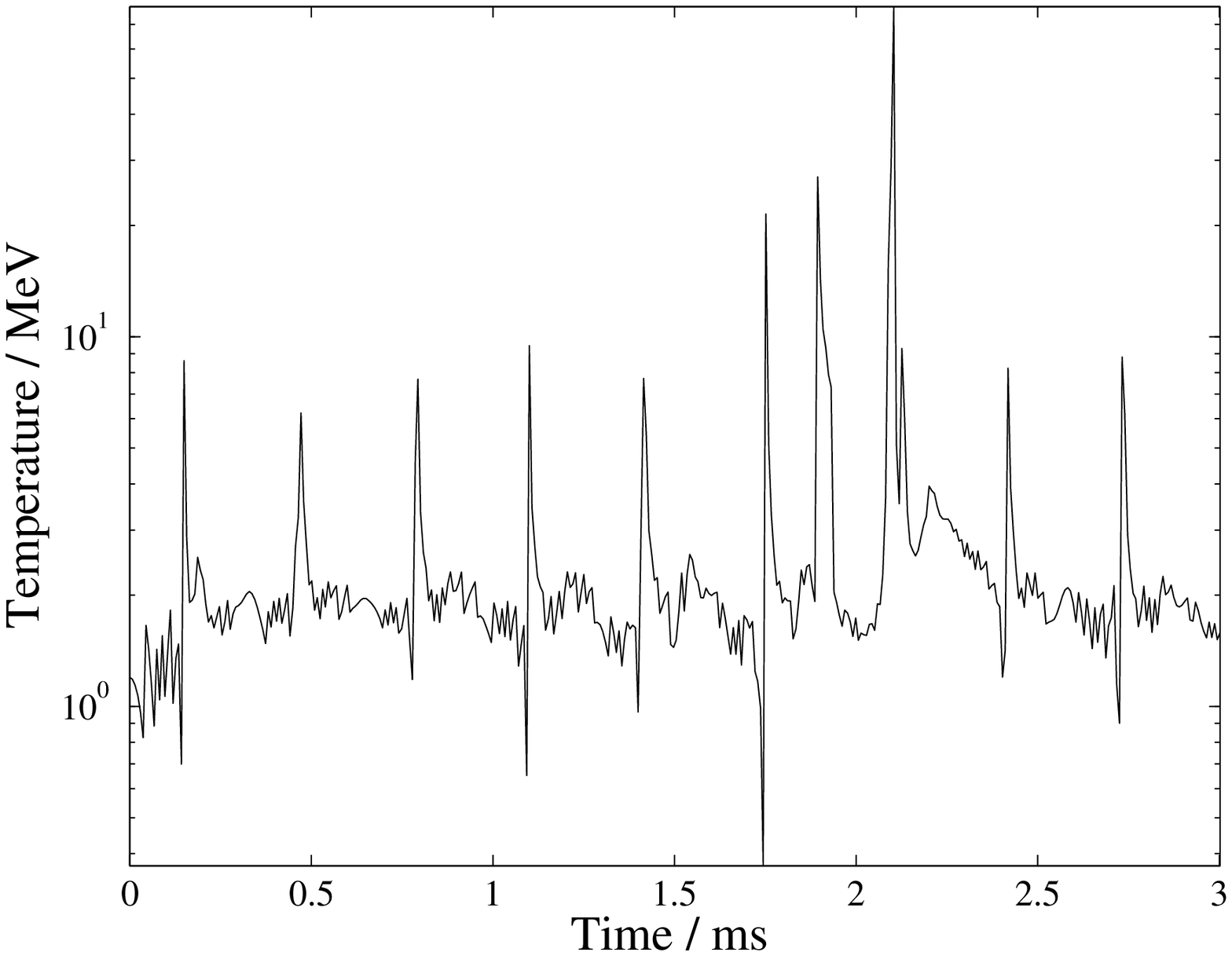}{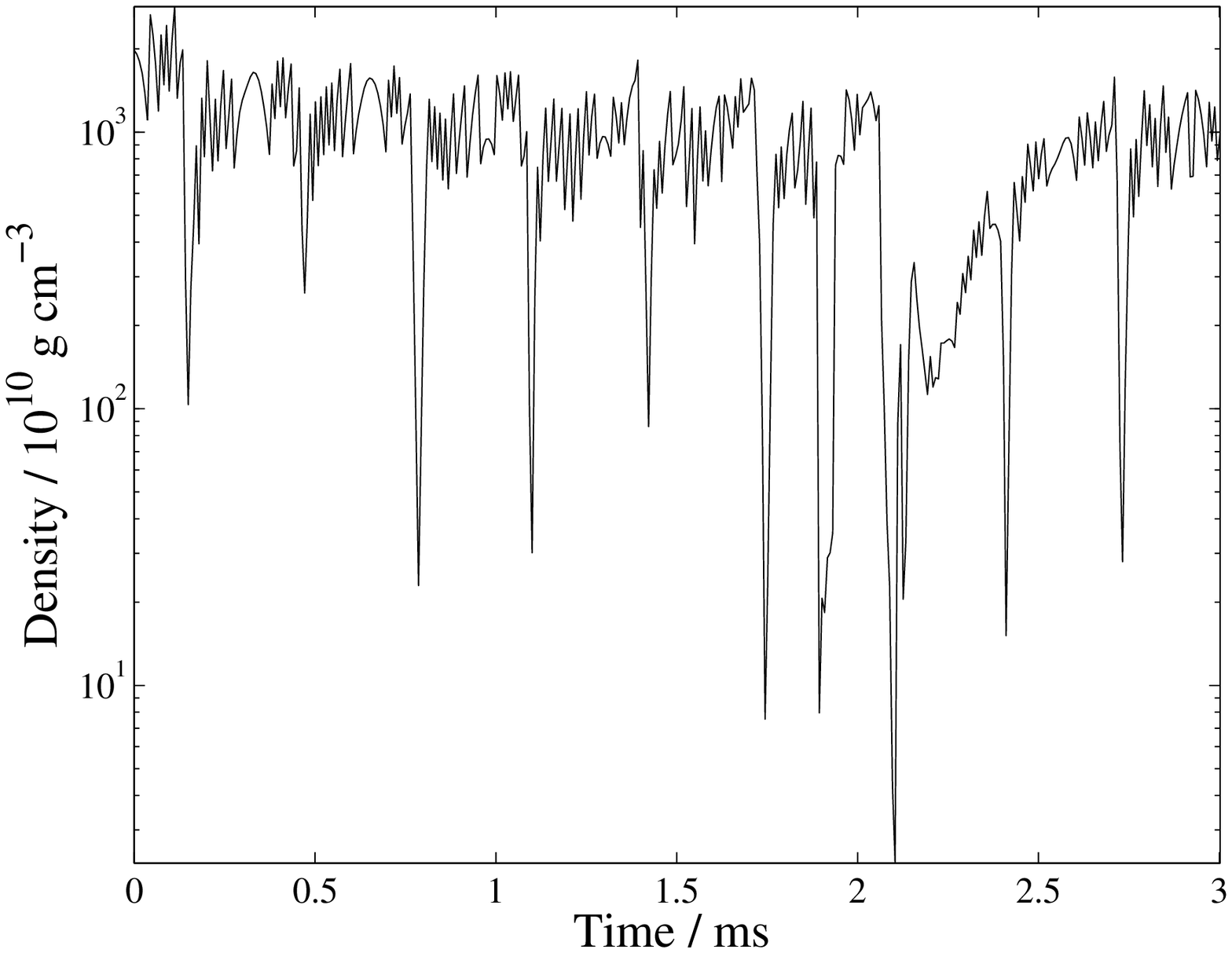}
\caption{{\itshape Left:\/} Neutrinosphere temperature versus time.
{\itshape Right:\/} Neutrinosphere density versus time. The initial
mass of neutron is 1.7M$_{\odot}$}
\end{figure}

\begin{figure}[!ht]
\centering \plotone{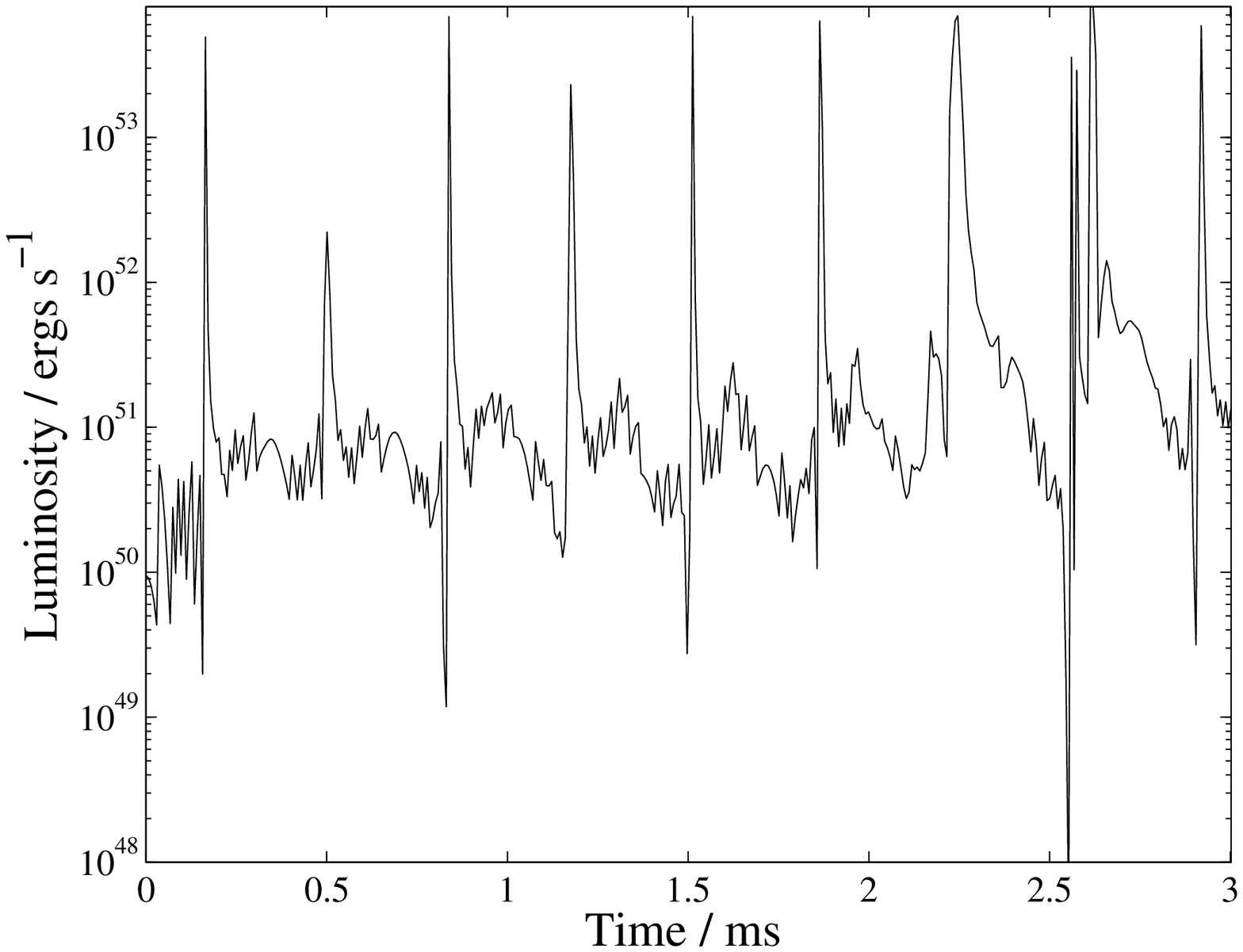} \caption{Neutrino luminosity as a
function of time of a star with 1.55M$_{\odot}$.}
\end{figure}

\begin{figure}[!ht]
\centering \plotone{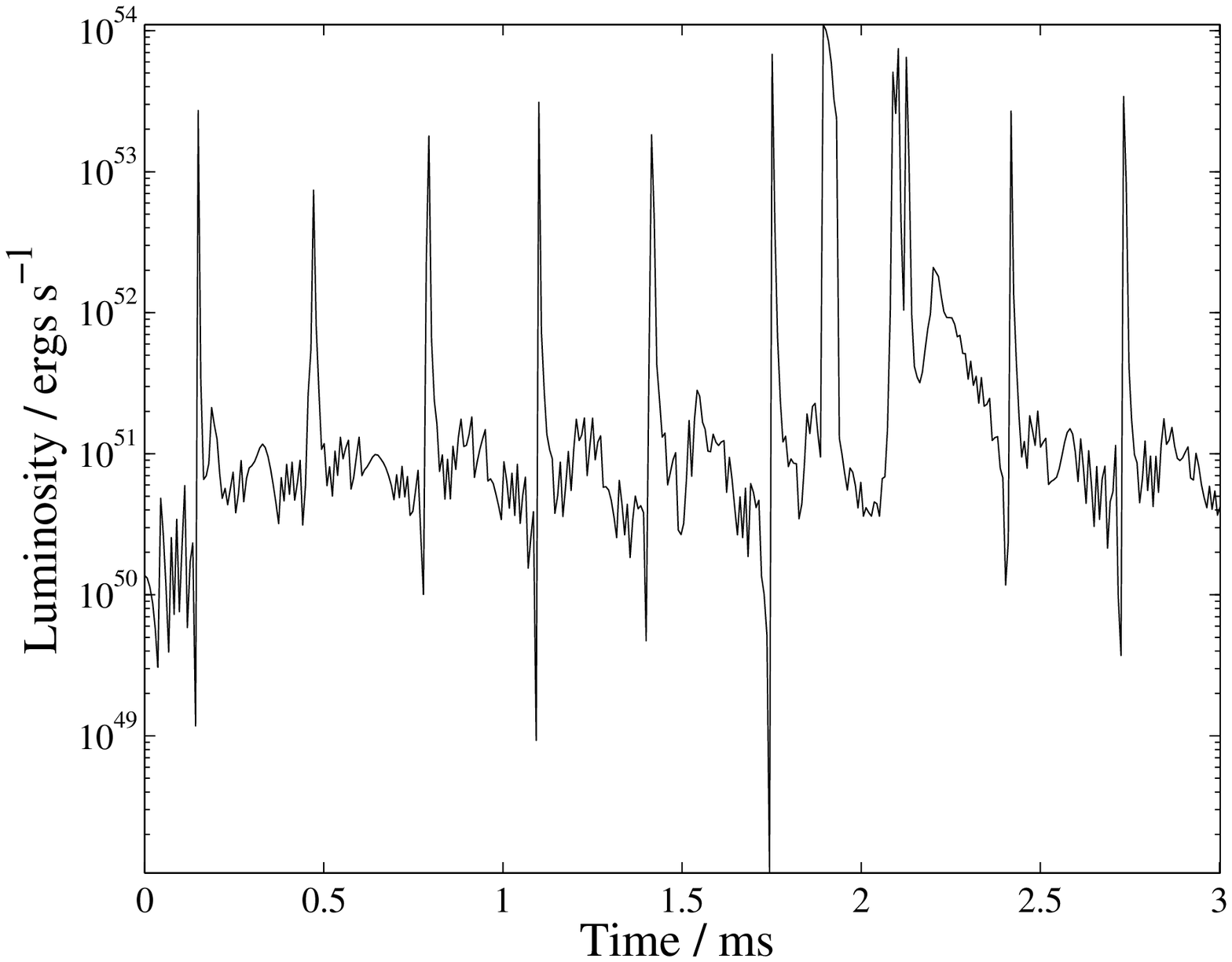} \caption{Neutrino luminosity as a
function of time of a star with 1.7M$_{\odot}$.}
\end{figure}

\vspace{0.2in}
\begin{figure}
\plotone{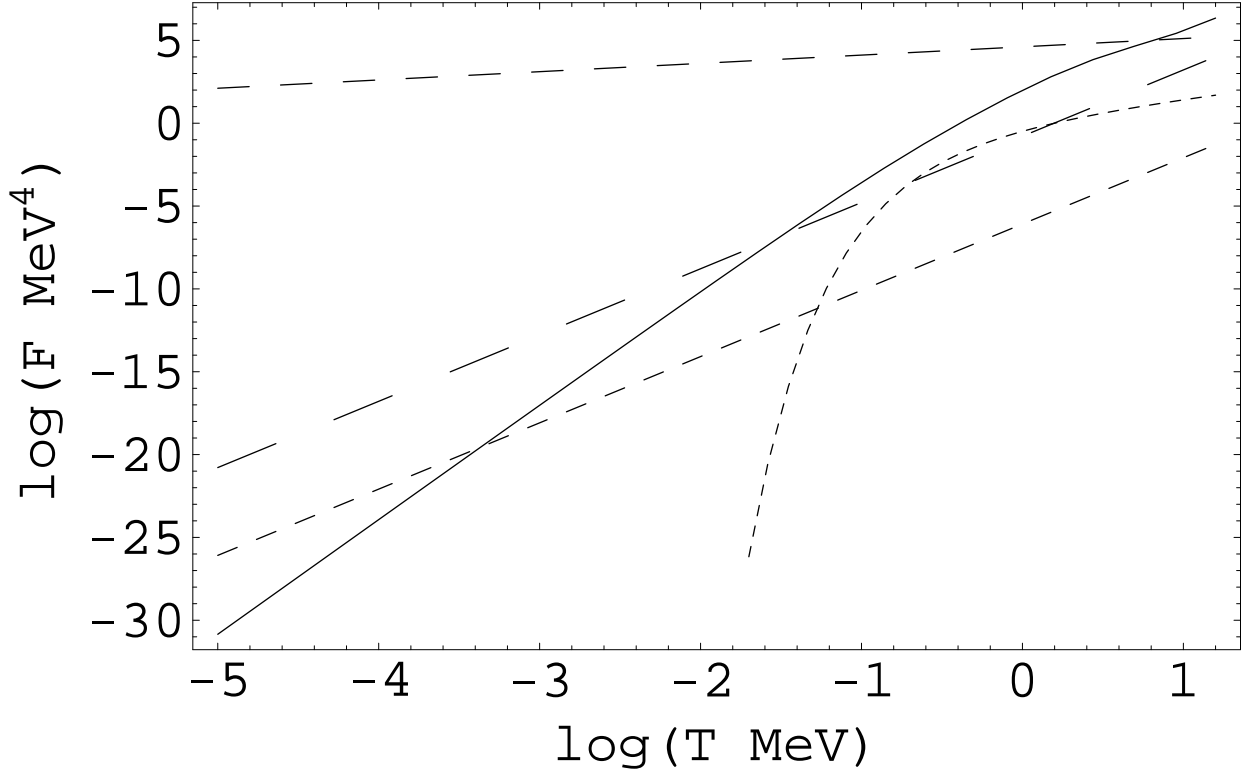} \caption{ Temperature variation of the energy
fluxes (in a logarithmic scale), emitted by the bare quark star
surface, via different radiation mechanisms: electron-electron
bremsstrahlung flux $F_{Br}^{(ee)}$ (solid curve), electron-positron
pair creation energy flux $F_{\pm}$ (dotted curve), quark-quark
bremsstrahlung energy flux $F_{q-q}$ (dashed curve), energy flux
$F_{\pi }$ due to pion emission (long dashed curve) and black body
radiation energy flux $F_{bb}$ (ultra-long dashed curve). For the
surface electrostatic potential of the quark star we have chosen the
typical value $V_I=14$ MeV, while the thickness of the electron
layer was taken $d=1000$ fm. For the electron-positron energy flux
the Fermi energy of the electrons $\varepsilon _F=18$
MeV$=$constant.
 For the energy density of the pion field at the
strange star surface we have adopted the value $\rho _{\pi }\approx
3.417\times 10^{5}$ MeV$^{4}$ $\approx 7.1\times 10^{31}$
erg/cm$^{3}$.
  }
\label{FIG5}
\end{figure}

\vspace{0.2in}
\begin{figure}
\plotone{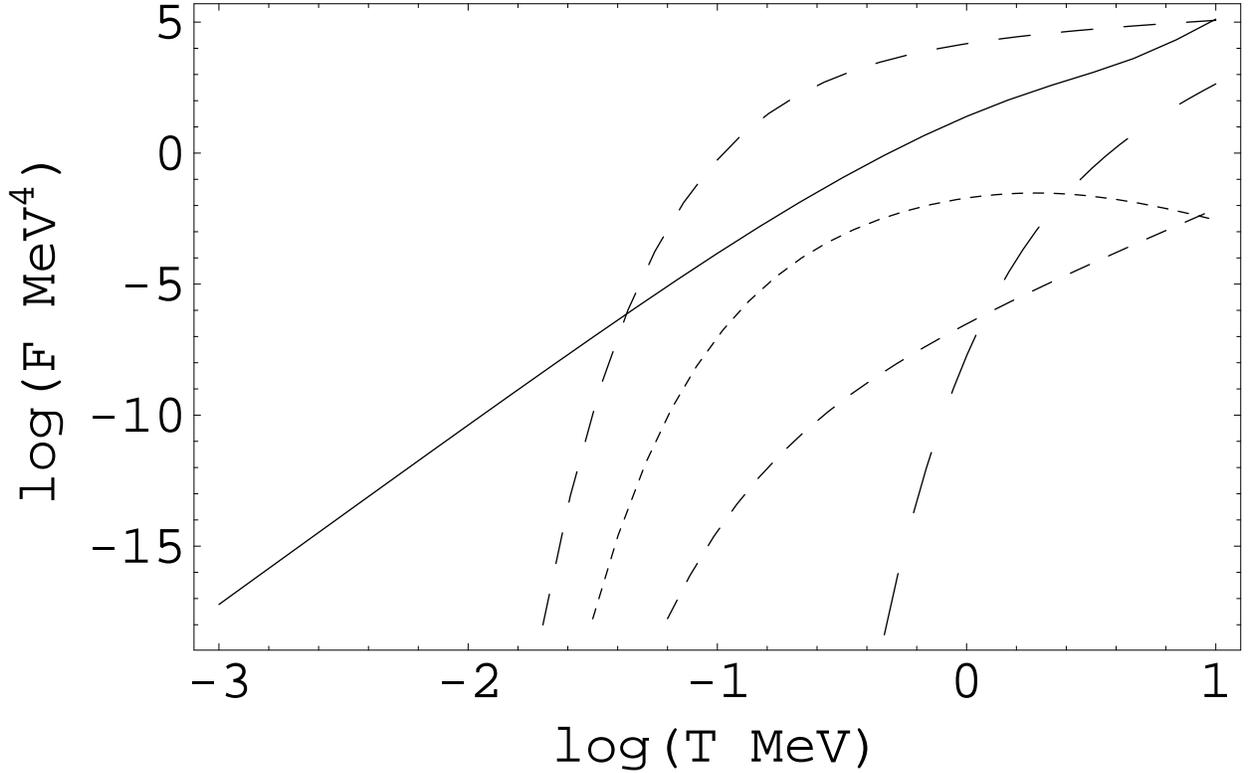} \caption{Temperature variation of the energy
fluxes (in a logarithmic scale), emitted by the superfluid bare
quark star surface, via different radiation mechanisms:
electron-electron bremsstrahlung flux $F_{Br}^{(ee)}$ (solid curve),
electron-positron pair creation energy flux $\left\langle F_{\pm
}\right\rangle $ (dotted curve), quark-quark bremsstrahlung energy
flux $F_{q-q}^{(\sup )}$ (dashed curve), energy flux $F_{\pi
}^{(\sup )}$ due to pion emission (long dashed curve) and the
thermal photon equilibrium radiation energy flux $F_{eq}$
(ultra-long dashed curve). For the surface electrostatic potential
of the quark star we have chosen the typical value $V$ thickness of
the electron layer was taken $d=1000$ fm. The energy gap $\Delta =1$
MeV.} \label{FIG6}
\end{figure}

\vspace{0.2in}
\begin{figure}[!ht]
\centering \plotone{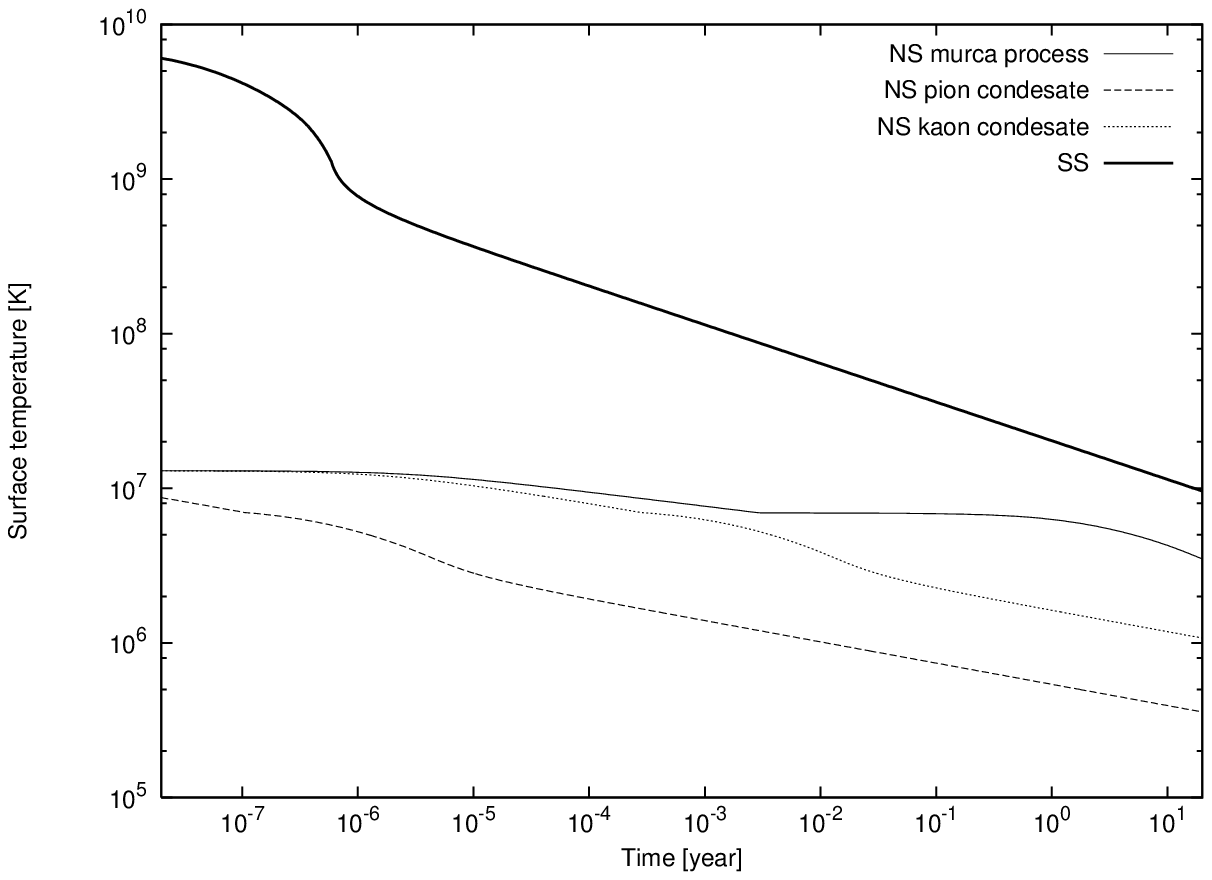} \caption{Surface temperature as a
function of time. The heavy solid line is the bare strange star
cooling curve. The light solid line, the dotted line and the dashed
line are the cooling curves of neutron star with modified URCA, kaon
condensate and pion condensate respective.}
\end{figure}

\vspace{0.2in}
\begin{figure}[!ht]
\centering \plotone{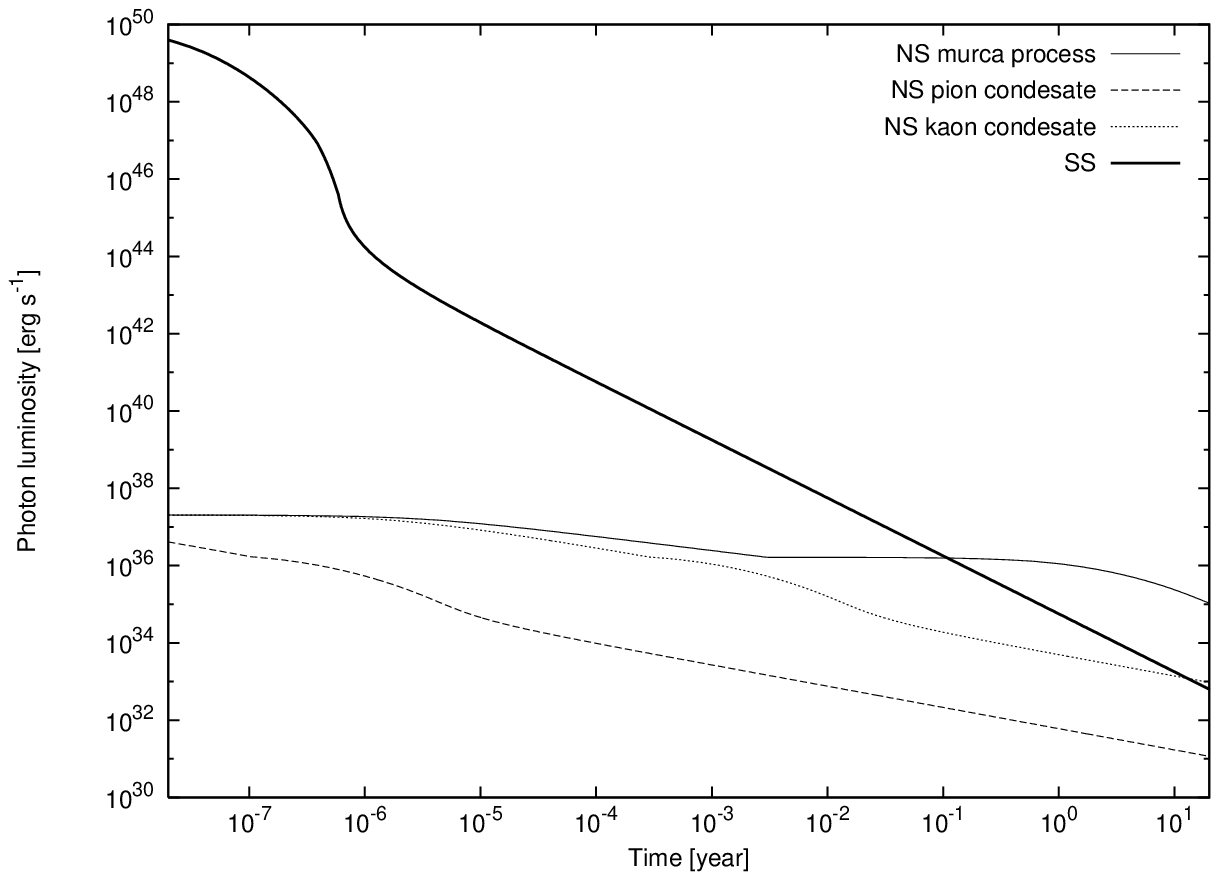} \caption{The surface photon luminosity
as a function of time. The heavy solid line is the bare strange star
cooling curve. The light solid line, the dotted line and the dashed
line are the cooling curves of neutron star with modified URCA, kaon
condensate and pion condensate respective.}
\end{figure}

\vspace{0.2in}
\begin{figure}[!ht]
\centering \label{} \plottwo{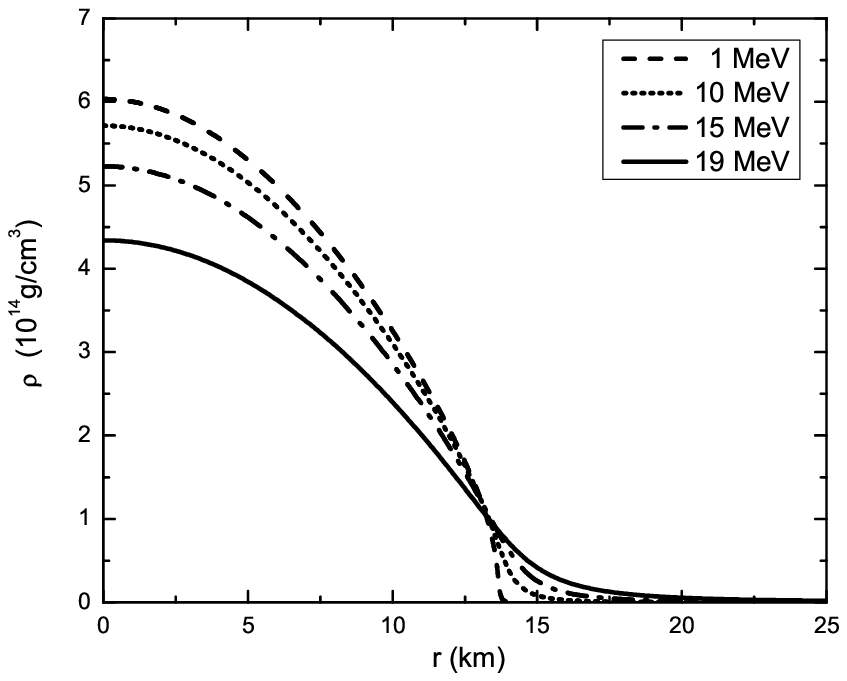}{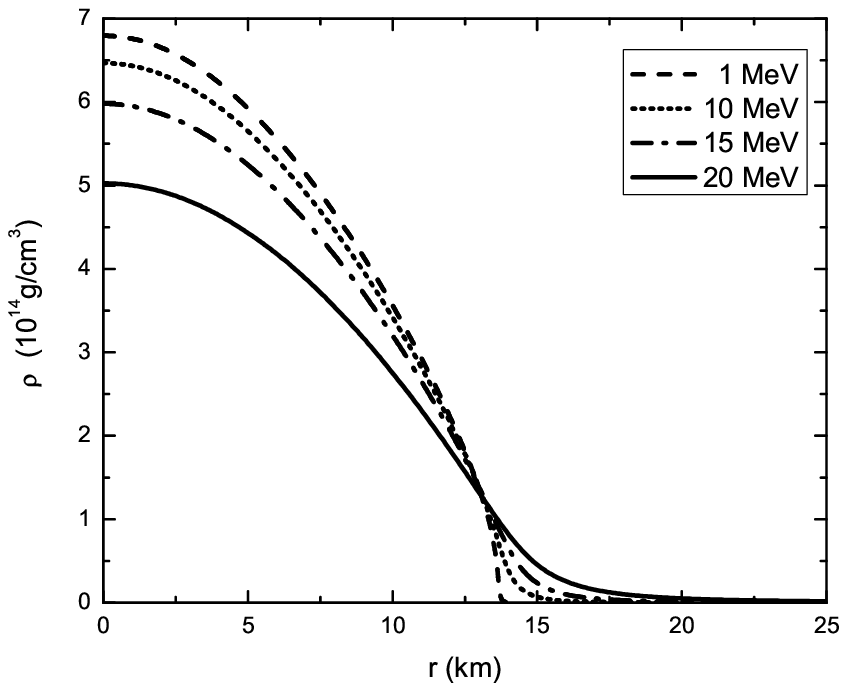}
\caption{{\itshape Left:\/} The energy density profile of stars with
mass 1.55 $M_\odot$ at different surface temperatures $T_s
=1,~10,~15,~19$ MeV. The central temperatures of the stars are $T_c
= 1.49,~14.7,~21.5,~26.0$ MeV respectively. {\itshape Right:\/} The
energy density profile of stars with mass 1.7 $M_\odot$ at different
surface temperatures $T_s = 1,~10,~15,~20$ MeV. The central
temperatures of the stars are $T_c = 1.57,~15.5,~22.8,~28.9 MeV$
respectively.}
\end{figure}
\end{document}